\documentclass[sigconf,nonacm]{acmart}

\AtBeginDocument{%
  \providecommand\BibTeX{{%
    \normalfont B\kern-0.5em{\scshape i\kern-0.25em b}\kern-0.8em\TeX}}}

\usepackage{_macros}
\usepackage{tikz}
\usepackage{soul}
\usepackage{xcolor}
\usepackage[table, dvipsnames]{xcolor} 
\usepackage{subfigure}
\usepackage{colortbl}

\usepackage[T1]{fontenc}
\usepackage[utf8]{inputenc}

\usepackage[most]{tcolorbox}

\definecolor{lightgray}{HTML}{bbbbbb}

\newcommand{\name}{NoteEx}

\begin{document}

\title[\name]{NoteEx: Interactive Visual Context Manipulation for LLM-Assisted Exploratory Data Analysis in Computational Notebooks}

\author{Mohammad Hasan Payandeh}
\email{mpayandeh@uwaterloo.ca}
\orcid{0000-0001-7712-3701} 
\affiliation{
  \institution{School of Computer Science\\University of Waterloo}
  \city{Waterloo}
  \state{Ontario}
  \country{Canada}
}

\author{Lin-Ping Yuan}
\email{yuanlp@cse.ust.hk}
\orcid{0000-0001-6268-1583} 
\affiliation{
  \institution{The Hong Kong University of Science and Technology}
  \city{Hong Kong}
  \country{China}
}

\author{Jian Zhao}
\email{jianzhao@uwaterloo.ca}
\orcid{0000-0001-5008-4319} 
\affiliation{
  \institution{School of Computer Science\\University of Waterloo}
  \city{Waterloo}
  \state{Ontario}
  \country{Canada}
}

\begin{abstract}
Computational notebooks have become popular for Exploratory Data Analysis (EDA), augmented by LLM-based code generation and result interpretation. 
Effective LLM assistance hinges on selecting informative context---the minimal set of cells whose code, data, or outputs suffice to answer a prompt. 
As notebooks grow long and messy, users can lose track of the mental model of their analysis. 
They thus fail to curate appropriate contexts for LLM tasks, causing frustration and tedious prompt engineering. 
We conducted a formative study (n=6) that surfaced challenges in LLM context selection and mental model maintenance. 
Therefore, we introduce \name{}, a JupyterLab extension that provides a semantic visualization of the EDA workflow, allowing analysts to externalize their mental model, specify analysis dependencies, and enable interactive selection of task-relevant contexts for LLMs. 
A user study (n=12) against a baseline shows that \name{} improved mental model retention and context selection, leading to more accurate and relevant LLM responses.
\end{abstract}

\definecolor{mycyan}{HTML}{FFE4B5}


\begin{teaserfigure}
    \includegraphics[width=1\textwidth]{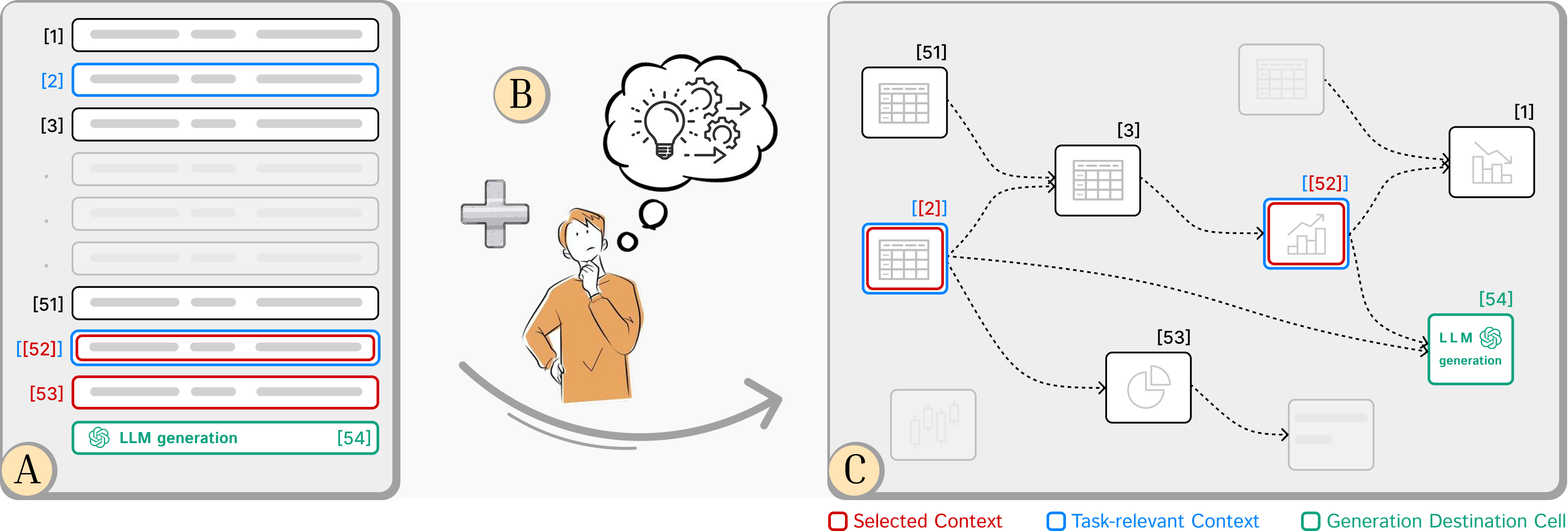}
    \vspace{-7mm}
    \caption{(A) Incorrect context selection: During Exploratory Data Analysis, computational notebooks often become long and messy. When an analyst uses LLMs to generate code in the green Cell \#54, automatic context selection methods often fail to select the correct task-relevant context. Instead of selecting the blue Cell \#2 and \#52, which are related to the analyst's mental model, the red Cell \#52 and \#53 are selected. 
    (B) Analyst's mental model: The problem roots in the complex dependencies existed in analysts' mental models, which represent how different notebook cells are interconnected based on the analyst's ideas and its hard to represent them using linear notebooks. Guided by their evolving mental model, analysts can externalize the complex dependencies with \name{}. (C) Proper context curation: \name{} allows the analyst to maintain an overview of their evolving mental model using its flowchart-style visualization, and to specify the proper blue cells as context for LLMs, augmenting automatic context selection. This leads to more accurate and relevant LLM responses to their task prompts. }
    \label{fig:system}
\end{teaserfigure}

\maketitle

\section{Introduction}

Exploratory Data Analysis (EDA) is an iterative and exploratory process in which analysts utilize data transformations, statistical techniques, and visualizations to understand datasets, identify patterns, and guide further decisions~\cite{tukey1977eda, Rule2018}.
Computational notebooks (\eg, Jupyter Notebooks~\cite{Perkel2018}) have become increasingly popular for conducting EDA, due to their convenience of writing and executing code segments (as code cells) dynamically, as well as examining outputs immediately and providing documentation (as markdown cells)~\cite{Rule2018, Lau2020, MOON2023, hara2015, EDAssistant2023}.
The rise of LLM-based tools (e.g., Copilot~\cite{github_copilot2023}) has transformed EDA with notebooks by assisting with tasks such as code generation for data processing and text generation for documentation~\cite{grotov2024untanglingknotsleveragingllm, weng2024insightlensaugmentingllmpowereddata}. 
However, effective LLM assistance hinges on providing an informative \textit{task-relevant context}~\cite{Kabongo2024contextselection}, which is a subset of notebook cells whose code, data, or output is necessary and sufficient to produce a high-quality response to the analyst’s prompt. 
Inappropriate contexts may lead to an increased risk of LLM ``hallucination'' and reduced quality in LLM responses~\cite{firooz2024lostindistanceimpactcontextualproximity, ning2024userllmefficientllmcontextualization, an2024makellmfullyutilize, WaitGPT2024}.

Analysts currently rely on either automatic or manual context selection to specify task-relevant cells~\cite{github_copilot2023, vscode}. 
In automatic selection, LLMs identify context based on explicit code dependencies, such as retrieving nearby cells that define variables or functions used in the current task. 
Selecting the context purely based on code dependencies may be imprecise, as it overlooks the analyst’s mental model (i.e., an analyst's internal understanding of their analysis workflow) underlying the code~\cite{Kabongo2024contextselection, github_copilot2023}. 
For instance, an analyst experimenting with alternatives might create nearly identical data-loading cells with slight modifications (e.g., different filters) using the same variable names, making it difficult for the LLM to select the correct cell. 
Therefore, such scenarios rely on the analyst’s underlying mental model to accurately determine which cells align with their current focus. To compensate, analysts turn to manual context selection, locating cells, recalling relationships, and curating context for each task. 
While providing more control, this manual process becomes increasingly time-consuming and cognitively demanding when the notebook grows long and messy~\cite{Head2019, chattopadhyay2023makemakesenseunderstanding, Chattopadhyay2020}.

In summary, both automatic and manual approaches fall short of providing task-relevant contexts effectively. 
Central to the issues are the various \textit{dependencies} existed in mental models, which represent how different parts of an analysis workflow (\ie cells in notebooks) are interconnected based on the analyst’s ideas. 
Some examples of such dependencies include sequential exploration (next step), parallel exploration (alternate paths), hierarchical stages (parent to child), fork-and-join (diverging or converging paths), and documentation~\cite{li2025composingdatastoriesmeta, ramasamy2023visualising, suh2023grammarhypothesesvisualizationdata}. 
Unlike code dependencies, mental-model dependencies are diverse, personalized, and non-linear~\cite{li2025composingdatastoriesmeta, ramasamy2023visualising}. 
To help manage and track mental models, prior work has explored strategies such as freely organizing notebook cells on a canvas~\cite{StickyLand2022, Sage3Harden2023, 2020-b2} or arranging them as hierarchical sublists~\cite{Harden2023, HardenPhD2023, Harden2022Exploring_Organization}. 
However, these approaches have not specifically addressed the unique challenges analysts face in context selection.

This work aims to support task-relevant context selection for LLM generation in EDA by incorporating mental-model dependencies.
We first conducted a formative study with six participants, in which they performed an EDA in VS Code using GitHub Copilot Chat for LLM assistance. 
The results reveal four recurring challenges: 
(CH1) a user's mental model evolves as the EDA progresses; 
(CH2) cell-specific metadata (\eg, execution status, order, outputs, and errors) are helpful, but impossible to remember and retrieve; 
(CH3) data variables are implicit, making them difficult to find and use in contexts; 
(CH4) pure manual context curation is burdensome, while automatic selection often fails.

To address the above challenges, we propose \name{}, a JupyterLab extension that allows analysts to interactively and visually manage their mental models and select appropriate context for LLM tasks during EDA.
\name{} provides a comprehensive view of the analyst's mental with a flowchart-style visualization, displaying cells as nodes and their dependencies as links (CH1).
The analyst can specify necessary mental model dependencies as links to augment the traditional automatic code dependency based context selection for more precise and higher quality LLM responses. 
The analyst only needs to create a link where they feel fit to clarify their mental models to LLMs.
Each node also contains information about its corresponding cell, including its position in the 1D notebook, content snippet, execution status, and output preview (CH2).
This is supplemented with a high-level overview on the top of the canvas, including a list of data variables along with the specific cells where they are defined and used (CH3).
Before prompting LLM, \name{} suggests task-relevant cells and data variables inferred from the specified mental model dependencies in an editable preview; analysts can add or remove items to finalize the context sent to the LLM (CH4).
Together, these novel features help analysts recall their mental models, identify task-relevant context, and select them for LLM assistance, effectively achieving various tasks such as code explanation, generation, modification, debugging, and interpretation of outputs.
    
To evaluate \name{}, we conducted a user study with 12 participants experienced in using computational notebooks and LLMs for EDA, comparing our system against a baseline.
The results indicate that \name{} significantly enhanced participants' EDA workflows by supporting effective mental model maintenance, notebook navigation, and LLM context selection, leading to more positive experience of using LLMs.
In summary, our paper makes the following contributions:
\begin{itemize}
    \item An interactive tool for EDA in computational notebooks, \name, which help users manage and externalize their mental models during the analysis with various informative supports, promoting more effective context selection for LLM-assisted tasks. 
    \item Empirical knowledge from a formative study on exploring the challenges in performing EDA with LLM in notebooks, and a user study on assessing \name{} that highlights the benefits of human-in-the-loop context selection, which would imply future tool design in a similar domain. 
\end{itemize}

\section{Related Work}

\subsection{Computational Notebooks and Exploratory Data Analysis}

Computational notebooks, such as Jupyter~\cite{Perkel2018}, Google Colab~\cite{Colab2019}, and Databricks~\cite{Databricks2019}, are interactive documents with cells containing text or executable code for computing results and generating visualizations. 
These cells are linearly arranged but can be reorganized, reshuffled, and executed in any order~\cite{Rule2018}. 
Because of this flexibility, users across various domains utilize computational notebooks to conduct Exploratory Data Analysis (EDA), an open-ended process including data transformation, statical inference, visualization, and so on~\cite{tukey1977eda}.
Data analysts use computational notebooks for tasks such as data processing, creating visualizations, and statistical analyses to uncover data trends~\cite{Rule2018}. 
Machine learning researchers prototype and experiment with models~\cite{Lau2020}. 
Educators employ notebooks for interactive lessons, live coding, and hands-on exercises with immediate feedback~\cite{MOON2023, hara2015}.

Due to the exploratory nature of EDA, notebook cells can be related based on an analysts' mental model. 
The mental model refers to the analyst’s internal, often non-linear, and personalized understanding of their analysis workflow, shaping how they interpret and structure their work~\cite{Chattopadhyay2020, christman20232d, Harden2022Exploring_Organization}. 
There are dependency relationship in the mental model, which describes how different pieces of the workflow (or cells) are related according to this mental model, transcending whether code or data actually flows between them~\cite{li2025composingdatastoriesmeta, ramasamy2023visualising, suh2023grammarhypothesesvisualizationdata}. 
Mental model dependencies are not the same as code dependencies, which strictly represent the output or state produced by one cell is required by another~\cite{ColinBrown2023, NBSearch2021}. 
Both matter in practice, but they serve different purposes: mental model dependencies communicate intent, rationale, alternatives, or narrative structure, whereas code dependencies ensure that variables, data, and computations are available for correct execution~\cite{Harden2023, Rule2018, Head2019, MOON2023}. 
For example, an analyst creates two identical data loading cells that import the same dataset with slight modifications (e.g., different filters or preprocessing steps) using the same variable names; such cases frequently happen in EDA when the analyst wants to experiment with different alternatives and copy \& paste code. 
When the LLM is prompted to generate a visualization, it may struggle to choose the correct data cell as the context. 
Or, an analyst exploring different ways to visualize the same dataset may want the LLM to generate code for a new type of plot, but these two plots are not code dependent. 

Taken together, these cases show that context selection based solely on code dependencies cannot reliably identify the cells needed for high-quality LLM responses. Analysts’ intents, captured through mental-model dependencies, often determine which cells should guide an LLM’s output. In this work, we explore augmenting dependency-based selection with lightweight, user-specified mental-model links, allowing analysts to mark semantic relations and identify the relevant cells, variables, and outputs for a prompt.

\subsection{AI-assistants for Computational Notebooks}

Proper coding assistants integrated into computational notebooks are essential, especially for users with limited coding experience~\cite{Kozlikova2017, Chattopadhyay2020}. 
It is shown that Natural Language Interfaces (NLI) can benefit users with various coding tasks such as explaining, generating, modifying, and debugging code, reducing the need for extensive coding expertise; 
example tools include FlowSense~\cite{FlowSense2020}, NL4DV~\cite{NL4DV2021}, and Eviza~\cite{Eviza2016}.

The rise of Large Language Models (LLMs) like ChatGPT~\cite{chatgpt} has pushed this field forward, creating new possibilities for NLIs~\cite{zhang2024naturallanguageinterfacestabular, weber2024computationalconversationalnotebooks}. 
First, LLMs demonstrate superior performance in coding tasks compared to traditional NLIs, which have increasingly been adopted in research, such as Grotov \etal's system for code debugging in computational notebooks~\cite{grotov2024untanglingknotsleveragingllm}. 
Second, LLMs excel in reasoning and generating insights aligned with users' analytical or programming intents~\cite{weng2024insightlensaugmentingllmpowereddata, CoLadder}, which would also be valuable in computational notebooks for uncovering patterns in datasets and visualizations.

Context selection in computational notebooks is the choice of which parts of the environment an LLM should use to respond to a prompt.
The context needs to be task relevant, which can be represented by a subset of notebook cells whose code, data, or output is necessary and sufficient for an LLM to produce a high-quality response~\cite{Kabongo2024contextselection}. 
In notebooks, task-relevant context may include cells and data variables~\cite{berret2024icebergsensemakingprocessmodel, rule2018thesis, iodide}. 
Effective context selection matters crucially because including too little context risks omissions and errors, while including too much can distract the model and degrade answer quality~\cite{firooz2024lostindistanceimpactcontextualproximity, ning2024userllmefficientllmcontextualization, an2024makellmfullyutilize, WaitGPT2024}.

Current assistants use two strategies~\cite{github_copilot2023, vscode}: (1) automatic selection (e.g., Copilot) based on the coding environment (current file, open files, cross-cell dependencies), and (2) manual selection where analysts specify cells/variables. 
Automatic selection is convenient but often follows code/proximity signals rather than analysts’ mental-model dependencies; in EDA, conceptual relations (\eg, alternatives, hypotheses, documentation links, versions, thematic groupings) drive relevance. 
This mismatch raises hallucination risk and reduces accuracy~\cite{firooz2024lostindistanceimpactcontextualproximity, ning2024userllmefficientllmcontextualization, an2024makellmfullyutilize, WaitGPT2024}. 
While Copilot, a popular LLM-powered notebook tool, effectively assists coding and explanations, its default automatic context limits analysts’ ability to steer the model toward mental model dependencies~\cite{github_copilot2023, vscode}.

Manual selection offers precise control but is burdensome in large, messy notebooks: analysts must remember relevant cells/outputs, locate variable definitions/uses, and verify execution status and errors. 
Without UI support that surfaces mental model dependencies, variables, and cell metadata at scale, this process is slow and error-prone, revealing a gap in IDEs like VSCode. 
Moreover, Copilot-like tools typically consider only code as context; incorporating data variables, cell outputs, and error messages could improve contextual understanding and task relevance~\cite{berret2024icebergsensemakingprocessmodel}.

Due to the deficiency of each method, in our work, we explore how to augment the automatic context selection with minimal manual context selection for EDA with LLM assistance. \name{} is designed with this new approach through an interactive canvas allowing analysts to specify their mental model dependencies when necessary with visualization and informative support (\eg, cell-specific metadata and variables), which could result in easier and more precise task-relevant context selection. 

\subsection{Visualizations for Computational Notebooks}

The aforementioned issues for large and messy notebooks, can be related to the linear (1D) representation of computational notebooks~\cite{chattopadhyay2023makemakesenseunderstanding, Keelawat2023, EDAssistant2023}.
In particular, the 1D representation manifests in the following challenges.
First, messiness, which occurs when interdependent cells clutter the notebook, making it difficult to track tasks~\cite{Head2019, Harden2023}. 
Second, tedious navigation, where users waste time scrolling through long notebooks to find specific cells or outputs~\cite{Harden2023}. 
Third, performance issues arise during non-linear analyses, where unnecessarily re-running all time-intensive cells significantly slows down the workflow, even though only a subset of cells needs to be executed to achieve the desired outcome~\cite{Harden2023, Lau2020}. 
Lastly, errors can arise from executing cells out of order, such as forgetting to initialize variables, leading to incorrect results that are difficult to identify~\cite{MOON2023, StickyLand2022, Merino2022}.

These challenges highlight the need for innovative solutions to improve the user experience when performing EDA in computational notebooks. 
Existing research suggests that organizing EDA cells in a 2D space can provide a more clear overview of the mental model, addressing some of the challenges.
Harden \etal~\cite{Harden2023, Harden2022Exploring_Organization, christman20232d} highlights three 2D cell-structuring approaches: Linear Patterns with single-column variations, Multi-Column Patterns for organizing semantic chunks, and Workboard Patterns with grouped or flowchart-like layouts. 
However, all the patterns, except Directed Graphs, rely on rigid structures, which may not align with realistic scenarios where cells have unstructured dependencies---referred to as ``dependency hell''---where any cell may depend on any other cell~\cite{Chattopadhyay2020}. 
Other tools utilizing 2D canvases~\cite{StickyLand2022, Sage3Harden2023} also inherit shortcomings. B2~\cite{2020-b2} incorporates some contextual information, such as variable names and previews of cells' outputs, but does not visualize analysts' mental model or execution dependencies. 
Another attempt to visualize cell dependencies is through tree structures~\cite{WhatsNext2023}, however, it does not provide sufficient detail to help analysts understand the EDA workflow or utilize the contextual information. Additionally, the dependencies are generated based on user interactions rather than capturing the analyst's mental model.
Flowchart-style visualizations, which are essentially Directed Graphs~\cite{Harden2022Exploring_Organization}, offer a holistic view that effectively illustrates complex unstructured dependencies, as seen in tools like VisFlow\cite{VisFlow2017}, VisComposer\cite{VisComposer2018}, and VMetaFlow~\cite{VMetaFlow2022}, and can be used to address the ``dependency hell'' problem in computational notebooks.

However, the above 2D canvas-based notebooks and visualization tools are not properly designed for LLM context selection, lacking explicit, queryable representations and mechanisms to curate, scope, and prioritize relevant context for prompting. 
They also fall short in exposing enough information for analysts to maintain their mental models for LLM tasks in EDA, such as cell-specific context (e.g., partial outputs or variable peeks) and notebook-level context (\eg, datasets, variable lineage, and execution state).
Inspired by the flowchart-style visualizations, \name{} is thus designed to support EDA tasks in computational notebooks by showing relationships among cells based on the evolving mental model dependencies.

\section{\name{} Design}

In this section, we first describe a formative study that informed our system design, highlighting the challenges analysts face when using LLMs for EDA tasks in notebooks.
We then present an overview of \name{} and its key features, and describe how it was designed and developed based on the formative study results.

\subsection{Formative Study}
\label{formative_study}

While the literature has suggested strategies to mitigate the issues with long, messy notebooks ~\cite{Harden2023, HardenPhD2023, Harden2022Exploring_Organization,StickyLand2022, Sage3Harden2023, 2020-b2}, they have not specifically investigated what are the specific challenges of analysts for context selection in this situation.
We thus conducted a formative study to investigate the challenges analysts face when using LLMs for EDA tasks, particularly when selecting task-relevant context in notebooks.

\subsubsection{Study Setup}
We recruited six participants (three males and three females, aged 24–29), referred to as P1-6, from a local university, including one PostDoc, three PhDs, and two Master's students. 
All had prior experience with EDA using computational notebooks (4.8 years on average) and LLMs (2.2 years on average). 
The study was approved by our institutional research ethic office and was conducted in-person. 
Participants were tasked with using VS Code as a notebook environment, with Copilot enabled, to review the analysis and continue exploring the dataset by identifying patterns and generating insights through data processing and visualization.
Half of them was provided with a pre-prepared, long, and messy notebook containing an initial analysis of a marketing dataset to simulate such a situation where they need to understand other analysts' workflow and context selection becomes more challenging. 
The other half, worked from scratch to enable them to build their mental model and select context related to the analysis they had done.
We adopted this setup because the messy-notebook condition quickly surfaces scale- and clutter-induced context challenges within the study time, while the from-scratch condition lets us observe context selection grounded in participants’ own mental models without the confound of interpreting someone else’s workflow. 
We conducted observations while participants were performing the task and took notes when necessary. Participants were also instructed to think aloud during the task.
Following the task, a semi-structured interview was conducted to gather their feedback. 
Participants were asked guiding questions to reflect on their experience of the EDA process, specify the challenges they faced, and potential tool features they might require. 
Each participant was remunerated \$20 for their time and effort during a 90-minute study.

\subsubsection{Challenges}

We transcribed all interview sessions and conducted thematic analysis~\cite{smith_qualitative_2024, thematic} in three stages: familiarization, inductive coding, and theme development. 
In familiarization, we imported the audio transcripts and text observations into NVIVO~\cite{NVIVO2014} and familiarized ourselves with the data. 
During inductive coding, we identified 105 initial codes, which were then organized into 8 categories to facilitate the identification of core themes. 
In theme development, we synthesized these categories to pinpoint four core themes related to the challenges faced by participants.

\textbf{CH1: A user's mental model evolves as the EDA progresses, complicating the selection of proper contexts.} 
We found that as the analysis unfolded, users’ understanding shifted and expanded, and the notebook could become long and messy \cite{Head2019, chattopadhyay2023makemakesenseunderstanding, Chattopadhyay2020}. 
Thus, only a small slice of the notebook was visible, making it challenging to keep this evolving mental model aligned with what existed, in what order, and how pieces related. 
Participants demanded temporal structure and a global overview to re-anchor their understanding. \pqt{There is no information on the order these cells are introduced ... like a timeline.}{P5}
They also emphasized the need for an overview when only partial content is visible, \pqt{When I see the notebook, it only shows a part of it and the notebook is quite long, so sometimes I don’t know what information it has.}{P3}

\textbf{CH2: Cell-specific metadata is helpful for context selection, but impossible for users to remember and retrieve.} 
It was discovered that deciding what to include as context depended on cell-level details, such as their metadata (\eg, execution status, order, outputs, and errors). 
Such information was invaluable but hard for participants to recall or gather at scale. 
When the notebook was large and messy, checking execution state, order, outputs, and errors could be slow and error-prone. 
Participants mentioned clear visual cues that could expose this information without scrolling through the notebook: \pqt{If the Jupyter notebook grows larger, there aren't enough distinguishable visual cues [to identify the cells, without navigating the notebook].}{P6}

\textbf{CH3: Data variables are implicit, making them difficult to find and use in contexts.} 
In notebooks and other forms of programming, variables lack visual salience---they are difficult to identify quickly at a glance, track, and select as context, especially when their definitions and uses are dispersed~\cite{rule2018thesis, iodide}. 
We observed that participants often needed to reference variables in their prompts as context for the LLM (e.g., to generate a visualization). 
They scrolled through the notebook, reviewing multiple code cells to identify the data variable names required for their prompts.
Sometimes participants opened the CSV file to inspect headers and sample rows before returning to the chat to manually type the column names into the prompt. 
This back-and-forth process for rediscovering variable definitions and schema was time-consuming and introduced uncertainty about exact names. 
Participants explicitly asked to surface variables as first-class objects for context curation. \pqt{It would be nice if we can let the users specify a specific set of variables that are going to be displayed in the menu bar.}{P6}

\textbf{CH4: Pure manual context curation is burdensome, while automatic selection often fails.}
We observed that starting from scratch for every LLM query could lead to extra effort, and automatic selection based purely on code dependencies often missed the participants' intents. 
They demanded contexts that reflect mental-model dependencies, plus transparency and control of the selection. 
\pqt{If we know [LLM knows our mental model] that, for example, this cell is dependent on these 2 cells. wouldn't it be better to just automatically include them unless the participants says otherwise, so default would be selected.}{P2} 
\pqt{Maybe you could find out what’s the best default. Automatically include them and then let the user unselect.}{P4} 
They also asked to see exactly what the LLM receives: \pqt{[when interactively select context] it could be more obvious what goes into the LLM.}{P4}

\subsubsection{Design Guidelines}
From the above challenges and the study results, we derive the following design guidelines to inform our development of \name{}.

\textbf{DG1: Externalize the mental model and necessary dependencies.} 
The system should provide a persistent, glanceable overview that makes the analyst’s mental model tangible in long, messy notebooks. 
Ways could include visualizing how cells relate (based on mental-model dependencies), when they were introduced (timeline or ordering), and what role they play (e.g., code vs. narration). This overview should support re-anchoring after navigation and scale, enable quick traversal along dependency paths, and reduce drift by showing the global structure even when only a small portion of the notebook is visible. 

\textbf{DG2: Surface cell-specific metadata to inform context selection.} 
The system should consolidate essential cell-level context needed for selection decisions, execution status and order, error states, and output, directly in an overview with clear, distinguishable visual cues. It should also present these signals without requiring scrolling or opening cells, so that analysts can rapidly judge whether a cell is up-to-date, relevant, or needs re-running before inclusion as task-relevant context.

\textbf{DG3: Expose data variables to raise the awareness of context.} 
The system should elevate variables to first-class entities by listing them alongside cells with definition cells, use cells, and lightweight schema previews (e.g., type, shape, columns, sample values). Selecting variables directly as context should be enabled, by showing its defining/ consuming cells and resolving ambiguity when multiple alternatives define the same variable; thus, variable-level context can become discoverable and actionable.

\begin{figure*}[bt]

    \tikzset{mylabel/.style={
        draw=black, line width=0.7pt, 
        double=white, double distance=0.7pt, 
        circle, fill=mycyan, 
        inner sep=1pt, text=black
    }}

    \begin{tikzpicture}
        \node[anchor=south west, inner sep=0] (image) at (0,0) {
            \includegraphics[width=1\textwidth]{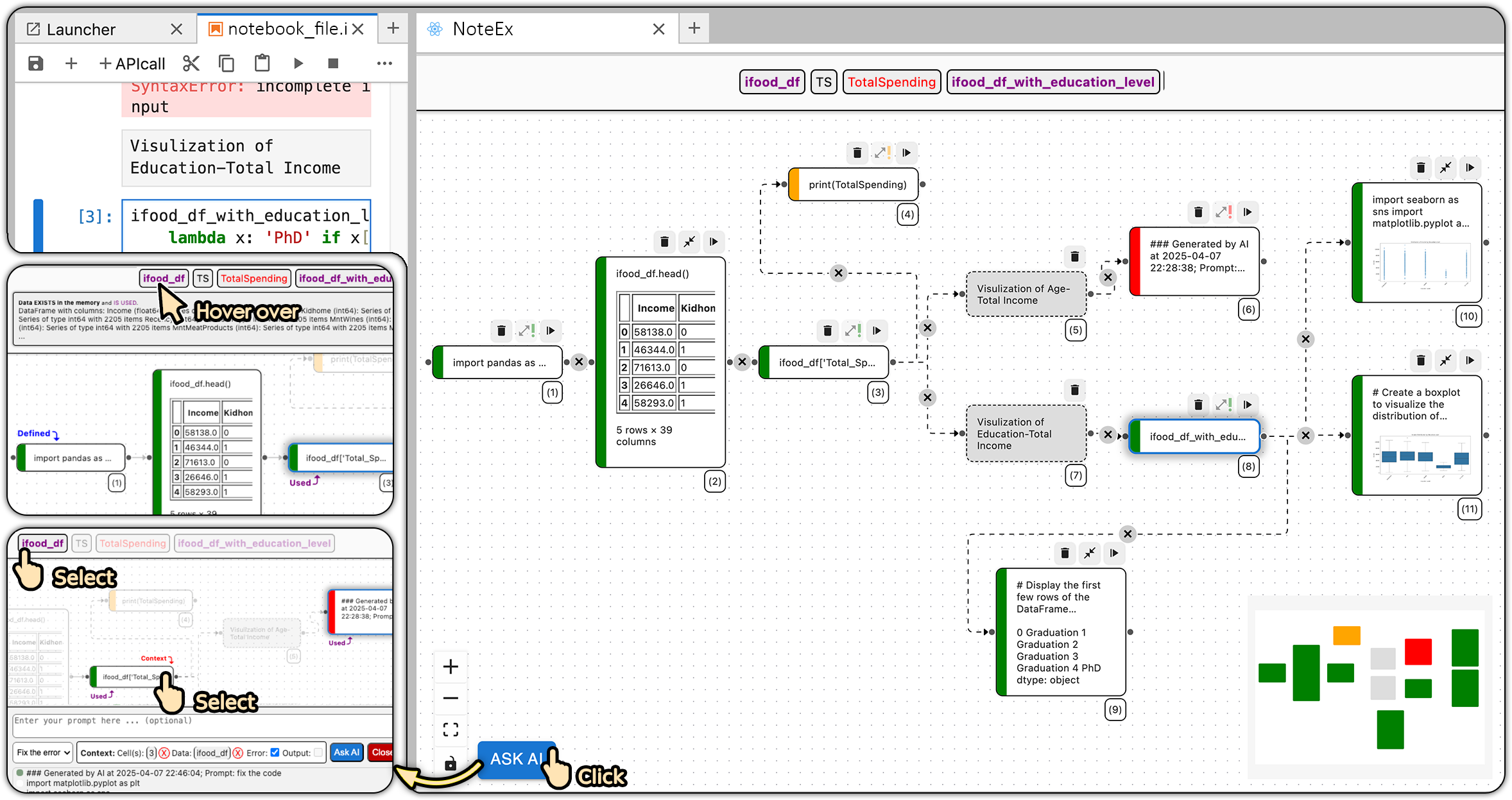}
        };

        
        \node[mylabel] at ([xshift=16pt, yshift=228pt]image.south west) {\textbf{A}};
        
        \node[mylabel] at ([xshift=151pt, yshift=218pt]image.south west) {\textbf{B}};
        
        \node[mylabel] at ([xshift=233pt, yshift=241pt]image.south west) {\textbf{C}};
        
        \node[mylabel] at ([xshift=14pt, yshift=139pt]image.south west) {\textbf{\scriptsize C1}};
        
        \node[mylabel] at ([xshift=121pt, yshift=13pt]image.south west) {\textbf{D}};
        
    \end{tikzpicture}
    
    \vspace{-2mm} 
    
    \caption{\name{} augments the traditional JupyterLab environment with four key elements for LLM-assisted EDA. (A) The Notebook View (B) The Canvas View (C, C1) The Data Information View (D) The LLM Assistant View}
    \label{fig:system} 
\end{figure*}
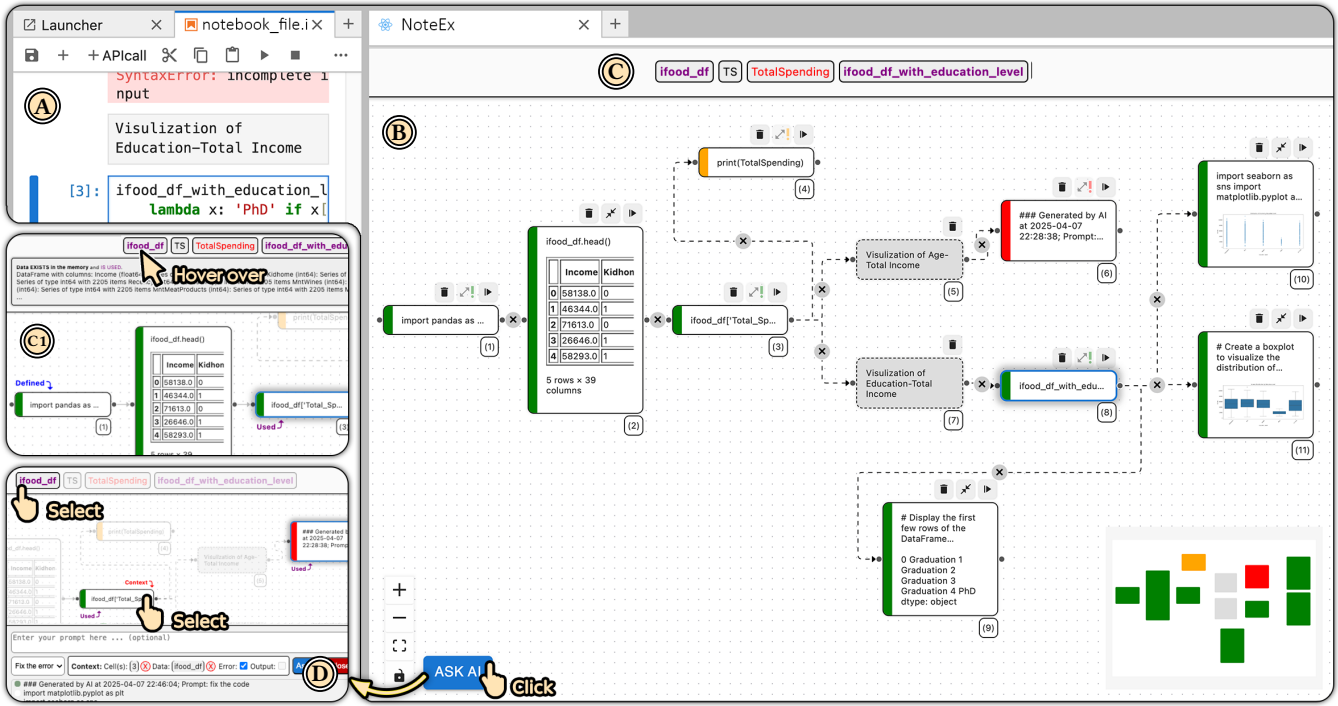

\textbf{DG4: Augment context selection with effective manual intervention.}
Based on the existing automatic context selection mechanisms, the system should allow analysts to add information based on their mental model to reflect the dependencies in their minds. 
Moreover, the system should offer a minimal, task-relevant ``suggested context'' based on the user’s mental-model dependencies.  
An interactive, editable preview that shows what will be sent to the LLM should also be provided, letting users add and remove contexts to override the default.
This reduces the time and likelihood of errors in context selection while maintaining full user control and clarity.

\subsection{\name{} Overview}

To operationalize the design guidelines (DG1–DG4), we developed \name{} as a JupyterLab extension, displayed as a tab alongside the Traditional Notebook View (Figure~\ref{fig:system}-A). 
The user interface of \name{} consists of three main views: Canvas View, Data Information View, and LLM-Assistant View.

The Canvas View (Figure~\ref{fig:system}-B) complements the Traditional Notebook View by presenting a flowchart-style overview of the analyst's mental model (DG1). 
It displays cells with notebook order, code/text snippets, output previews, variables defined or used, and execution statuses (DG2). It also enables users to connect cells to represent mental-model dependencies and externalize their mental models (DG1). 
In contrast, the Traditional Notebook View offers detailed information, such as full code/text and outputs. 
Both views are synchronized, so any change made in one view is instantly reflected in the other.
The Data Information View (Figure~\ref{fig:system}-C) displays a list of data variables, including variable names, structures/schemas, and the cells where they are defined or used (DG3). 
The LLM-Assistant View (Figure~\ref{fig:system}-D), accessible by clicking on the ``ASK AI'' button, provides assistance for coding and analysis tasks such as generation, modification, debugging, explanation, and output interpretation. 
Before a prompt is sent, the assistant shows a suggested set of task-relevant cells and data variables inferred from mental-model dependencies with an editable preview; 
users can add or remove contexts and specify the final set to be sent to the LLM (DG4). 
By using these views in \name{}, analysts can externalize their mental models, and leverage the specified dependencies in their minds as well as other visualized key information of cells to effectively assemble and pass task-specific context, at the level of both cells and variables, to the LLM for high-quality responses.


\section{Usage Scenario}
\label{usage_scenario}

Before introducing the detailed design and implementation of \name{}, we present a usage scenario that illustrates how the system supports context selection during LLM-assisted EDA and addresses the identified challenges.

\begin{figure*}[bt]
  \centering
  \includegraphics[width=\linewidth]{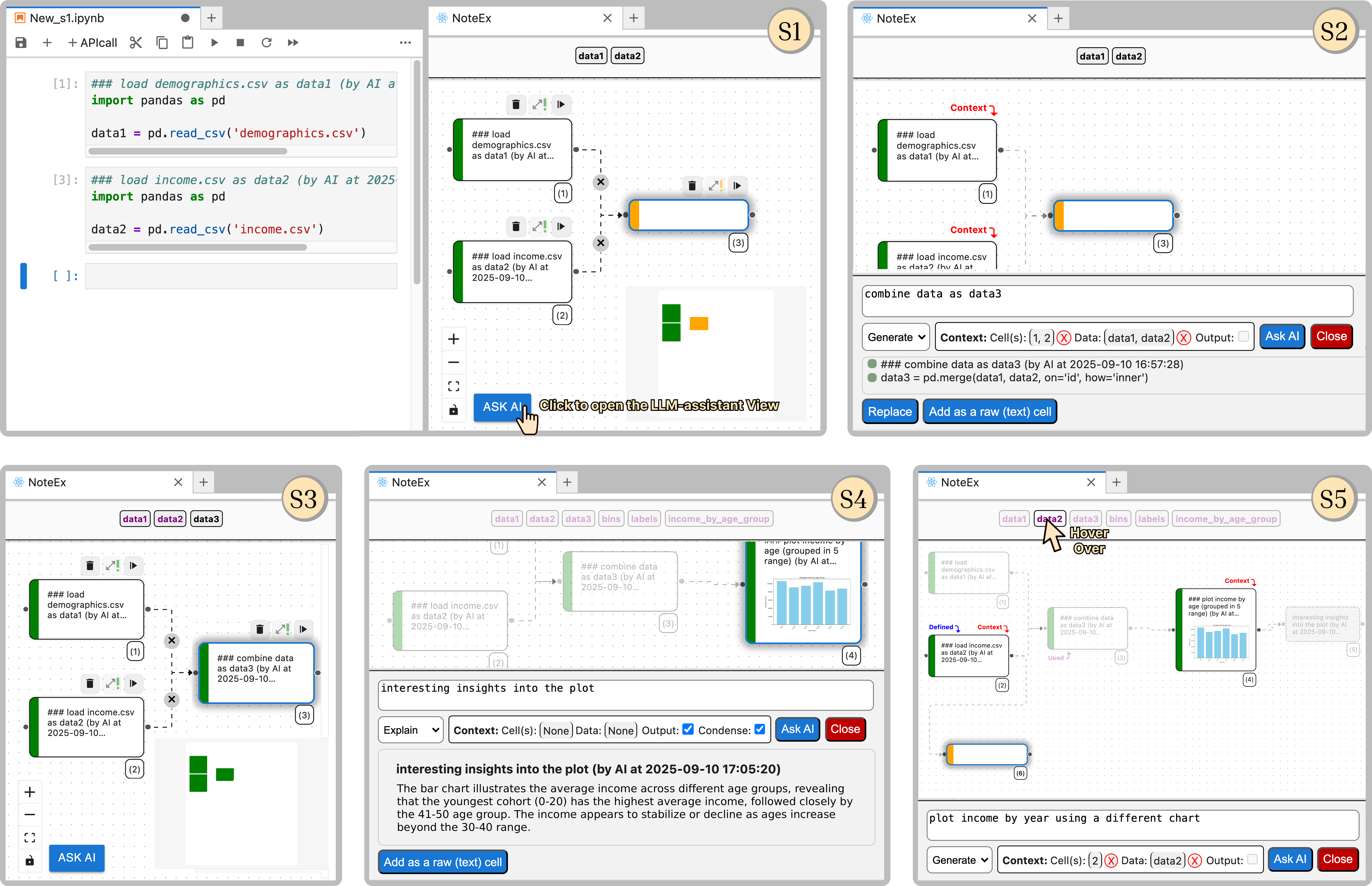}
  \vspace{-7mm}
  \caption{Key steps related to the usage scenario (Section~\ref{usage_scenario}), demonstrating how Sarah uses \name{} to perform her EDA tasks.}
  \label{fig:usage_scenario_1}
\end{figure*}

Sarah, a data scientist, is analyzing a marketing dataset to understand the relationship between customer demographics and income. She starts with an empty notebook in JupyterLab. 
In the Traditional Notebook View, she creates two data-loading cells (Figure~\ref{fig:usage_scenario_1}-S1; Nodes 1 and 2), one for demographics and one for income. 
Sarah wants to combine the two datasets to enable income-by-demographics analysis.
Thus, in the Canvas View, she drags a link from the first data-loading node (Node 1) into empty space. 
As a result, the system creates a new node linked to the data-loading node, which also appears as a cell in the Traditional Notebook View (Figure~\ref{fig:usage_scenario_1}-S1; Node 3).
She then connects the second data-loading node (Node 2) to the new node, capturing a fork-and-join pattern with a converging stage that mirrors her mental model of combining the data (CH1). 

To generate actual code, she selects Node 3 and opens the LLM-Assistant View by clicking on the ``ASK AI'' button (Figure~\ref{fig:usage_scenario_1}-S1). 
The assistant proposes a task-relevant context---both data-loading cells (Nodes 1 and 2) and their data variables (\texttt{data1} and \texttt{data2})---based on dependencies inferred from her externalized mental model (CH4; Figure~\ref{fig:usage_scenario_1}-S2). 
She writes the prompt ``combine data as data3'' and clicks ``Ask AI'' (Figure~\ref{fig:usage_scenario_1}-S2). 
The LLM generated code meets her needs, so she inserts it into cell (Node 3) by clicking ``Replace'' (Figure~\ref{fig:usage_scenario_1}-S2). 
Three data variables---\texttt{data1}, \texttt{data2}, and \texttt{data3}---are also automatically added to the Data Information View, corresponding to the demographics (Node 1), income (Node 2), and combined data (Node 3), respectively (CH3; Figure~\ref{fig:usage_scenario_1}-S3).
This immediately clarifies the exact variable names and provenance, allowing her later to select the dataset as context without scrolling through the notebook.

Next, she plans to make a plot showing income by age group based on the combined dataset.
Thus, from the Node 3, she creates another empty cell (Node 4) via dragging a link to the empty space on the canvas. 
To quickly surface interesting patterns, she turns to the LLM-Assistant View, switches the task type to ``Explain,'' keeps the ``output'' and ``condense'' options selected, and writes ``interesting insights into the plot'' to obtain a short summary of the plot (Figure~\ref{fig:usage_scenario_1}-S4). 
After reading the generated summary, she clicks ``Add as a raw (text) cell'' to insert the summary as a node (Figure~\ref{fig:usage_scenario_1}-S5; Node 5) to the Canvas View and as a cell to the Traditional Notebook View. 
This enables her to pick up from here later and supporting the maintenance of her evolving mental model (CH1).

Now she wants a second plot using only the income data. 
She creates an empty cell (Node 6) from the Node 2 
and opens the LLM-Assistant View. 
The default suggestions include the income-loading cell (Node 2) and the variable \texttt{data2} as context. 
To verify, she hovers over the variable in the Data Information View and sees a blue ``Defined'' label above Node 2, meaning it is defined in this node (Figure~\ref{fig:usage_scenario_1}-S5; CH4). 
This ensures only the relevant cells and variables are included, preventing the LLM from being misled by unrelated content. 
She also adds the first plot (Node 4) to the context to indicate she wants a different type of chart (Figure~\ref{fig:usage_scenario_1}-S5). 
The LLM generates code for a line chart, which she inserts to the Node 6 and execute the corresponding cell to produce the second plot.

\begin{figure*}[bt]
  \centering
  \includegraphics[width=\linewidth]{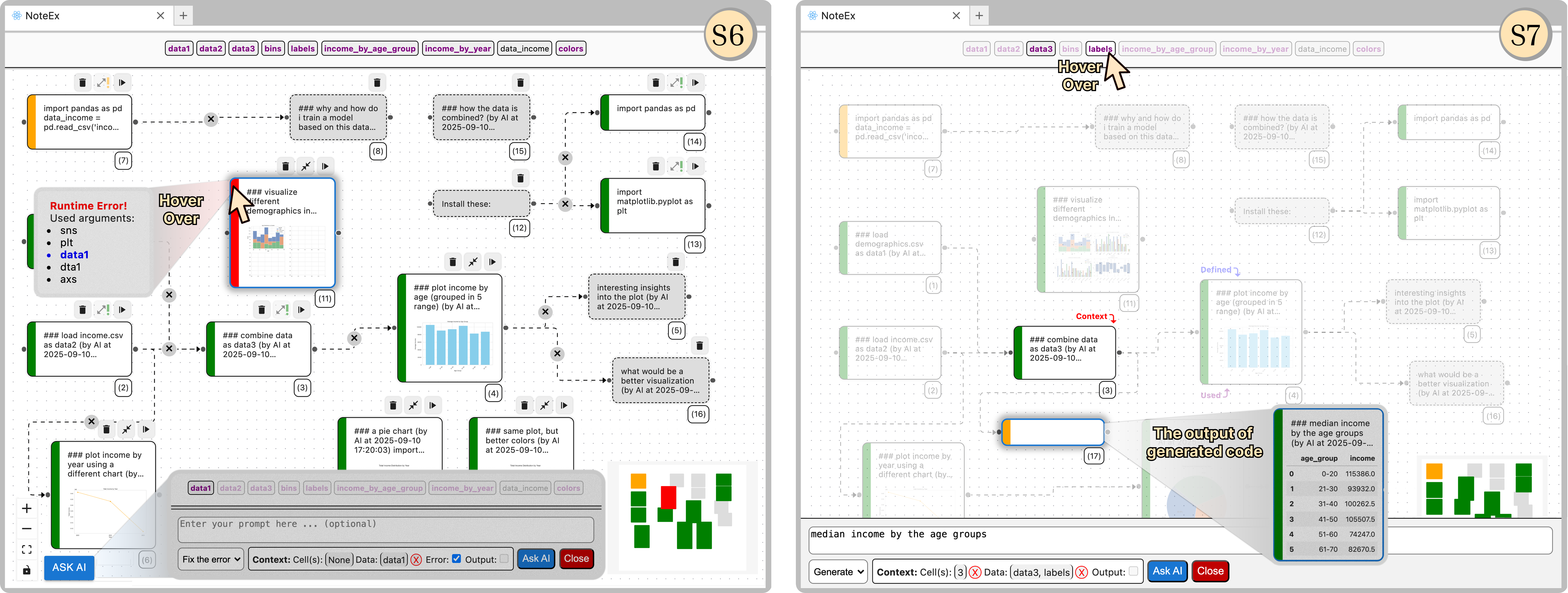}
  \vspace{-7mm}
  \caption{Key steps related to the usage scenario (Section~\ref{usage_scenario}), demonstrating how Sarah uses \name{} to perform her EDA tasks.}
  \label{fig:usage_scenario_2}
\end{figure*}

As Sarah continues, the notebook grows long (Figure~\ref{fig:usage_scenario_2}-S6). 
Despite the length and clutter in the Traditional Notebook View, the Canvas View provides a clear overview of her mental model (CH1), including dependencies and cell-specific metadata (CH2). 
The overview keeps her oriented and helps her navigate to the right stage of the workflow, while cell-level details (execution state, errors, output previews) help her decide which cells to include as context and which ones need fixing.

As shown in Figure~\ref{fig:usage_scenario_2}-S6, she creates only a few connections, just enough to clarify her mental model to the LLM generation. 
She notices Node 11 that is not connected to any others, marked with a red bar indicating it failed to execute. 
To fix it, she opens the LLM-Assistant View, where the task type is set to ``fix the error'' and the error checkbox is checked (Figure~\ref{fig:usage_scenario_2}-S6). 
Because the Node 11 is unconnected, she receives no suggestions. 
She thinks that used data variable in that cell might be relevant context. 
She hovers over the red bar to locate the data used by Node 11, sees that \texttt{data1} is used, and adds it as context (CH2; Figure~\ref{fig:usage_scenario_2}-S6). 
This helps the LLM have sufficient context and produce a high-quality fix.

Sarah suddenly has a thought on computing median income by age group, so she creates an empty cell (Figure~\ref{fig:usage_scenario_2}-S7; Node 17). 
Then, she connects Node 17 to Node 3, which is related to the combined data. 
Afterward, she opens the LLM-Assistant View, which suggests only Node 3 and its associated data variable (\texttt{data3}) as the necessary context. 
Since she wants to calculate the median using the same groups shown in Node 4, she needs to identify the variable that stores the age ranges and include it as context.  
To locate it, she hovers over the variables in the Data Information View (Figure~\ref{fig:usage_scenario_2}-S7). 
When she hovers over the ``labels'' variable, a ``defined'' label appears in the Canvas View above Node 4, indicating that this variable specifies the bins (CH3).  
She then selects the ``labels'' data variable as additional context (CH4). 
Finally, she generates the code for median income by age groups and inserts the code into the empty cell (Figure~\ref{fig:usage_scenario_2}-S7; The output of generated code).

Through the Canvas View for externalizing and maintaining her mental model, the Data Information View for variable-centric selection, and the LLM-Assistant View for suggestion-based, editable context sets, Sarah preserves her mental model as the notebook scales and consistently supplies task-relevant context to the LLM.

\section{\name{}}

JupyterLab provides a comprehensive environment that includes the JupyterLab interface, a Jupyter server, Notebook files, and a Python Kernel. 
\name{} is implemented as a custom extension built on top of the JupyterLab interface, developed using TypeScript~\cite{TypeScript} and React~\cite{react}. 
It appears as a dedicated tab alongside the linear cell view in the Notebook, automatically opening whenever a user creates or opens an \texttt{ipynb} file.
All data related to this extension is stored within the \texttt{ipynb} file’s metadata. 
This ensures that by simply saving the Notebook, any extension-specific data is also saved as JSON in the file, eliminating the need for a separate backend or database.
Communication between \name{} and the underlying notebook (or broader JupyterLab environment) leverages JupyterLab’s built-in extension APIs, which provide a robust interface for sending and receiving data.

In this section, we detail the key components of \name{}, including the Canvas View, Data Information View, and LLM-Assistant View.

\subsection{Canvas View}

The Canvas View of \name{} (Figure~\ref{fig:system}-B) includes a flowchart visualization of nodes and links, providing a overview of the analyst's mental model (DG1).
Each node in this view corresponds to a notebook cell, with several attributes displayed (DG2). 
First of all, nodes are labeled with the cell's order number. 
A cell with a white background and a color-coded bar on the left is a code cell, while a gray background without a bar indicates a raw or markdown cell.
The execution status of code cells is visually encoded using three colors: red indicates a runtime error, orange signifies that the cell has not been executed or its content changed, and green shows successful execution.
Moreover, each node contains a snippet of the cell’s content, including a portion of the code for code cells or text for raw and markdown cells. 
The nodes also display a preview of the output, which could be visualized as a plot or represented as text, such as printed data in Python. 
Further, nodes for the currently active cell is highlighted with a blue border and shadow (Figure~\ref{fig:system}-B-Node 8). 

Analysts can perform various actions, such as removing, expanding, and collapsing the node, via the buttons on it. 
They can also execute all cells along the path originating from the current node. 
Nodes can be freely moved within the canvas by dragging and dropping, and clicking on a node changes the active cell, automatically scrolling the Notebook View, as shown in Figure~\ref{fig:system}-A.
Hovering over the colored bar provides additional information about the cell’s execution status and the data variables used within it. 
Hovering over the play button displays the execution path, and clicking it executes all cells along the path.

Links connect nodes and represent mental model dependencies between cells.
They are visually represented as animated dashed arrows with distinct tails and heads to indicate the source and target. 
Each node is equipped with input and output handles to enable the creation of links between nodes. 
Analysts can establish a link by dragging from the output handle of one node and dropping it onto the input handle of another. 
Multiple input and output links for a single node is supported.
If a link is dropped onto an empty area of the canvas rather than onto another node’s input handle, a new node is automatically created. 
This new node is connected to the original node and is simultaneously set as the active node in a notebook. 
Links can also be removed by clicking the cross icon located at the midpoint of the link.

\subsection{Data Information View}

The Data Information View (Figure~\ref{fig:system}-C) offers detailed insights into the variables defined within the cells, presenting them in a list format (DG3). 
It complements both the Canvas View by enhancing overall mental model understanding and the LLM-Assistant View by aiding interactive context selection.
Each variable is visually coded with color to indicate its status. 
Red indicates data not in memory but being used, purple signifies data in memory and is used, and black represents data either in or out of memory but not being used.
Hovering over a variable name reveals its data structure, if it is loaded into memory. 
Additionally, this interaction highlights the locations of the cells in the Canvas View where the variable is defined and/or used. 
These locations are annotated with the text ``defined'' (in blue) or ``used'' (in purple), positioned above the relevant nodes (Figure~\ref{fig:system}).

\subsection{LLM-assistant View}

The LLM-assistant View (Figure~\ref{fig:system}-D) allows analysts to interact with LLM (GPT-4o-mini) to work with the active cell (Figure~\ref{fig:system}-B-Node 8). 
This view is accessed by selecting an active cell and clicking the ``ASK AI'' button. 
Built as an inline LLM, analysts can input prompts and optionally select portions of the cell's code or text to feed to the LLM (DG4). 
If a selection is made, the full code is provided as context, with the selected portion marked for change.

Analysts can customize the context for the LLM through several options. 
They can select the task type based on the active cell (e.g., code generation, modification, debugging, and explanation for code cells, or narration writing and editing for raw/markdown cells). 
Context cells are selected by clicking nodes in the Canvas View, and data is chosen by clicking variables in the Data Information View. 
Selected items are highlighted by changes in opacity. 
Additionally, checkboxes allow users to include runtime errors (when present) and cell outputs (e.g., images, text) as context for relevant tasks.

By default, the LLM-assistant View opens with a set of task-relevant context inferred from the analyst's externalized mental model dependencies. 
This includes cells directly linked to the active cell through incoming connections, along with the variables defined in those cells. 
These suggestions appear visually highlighted in both the Canvas View and the Data Information View, so that analysts can then verify and adjust the context by adding or removing cells and variables (DG4).

The prompts sent to the LLM API are dynamically generated based on the task type and selected context by analysts, ensuring that the responses are appropriate for the specific scenario (see Appendix~\ref{AppendixI}). 
Once a response is generated, it appears in a text box below the context selection box. 
If the task involves code modification, The lines with changed code are marked by a green circle at the start of each line. 
Analysts can freely edit the code before replacing the cell's content.

\begin{table*}[!htb]
\centering

\caption{Post-task questionnaire items (Q1--Q39) with test statistics and $p$-values, comparing user responses between Baseline and \name{}. All questions used a 7-point Likert scale (1 = strongly disagree, 7 = strongly agree), except for Q4--Q7, which were inverted (7 = strongly disagree, 1 = strongly agree). }
\label{tab:subj-questions}
\vspace{-3mm}
\renewcommand{\arraystretch}{1.2}
\setlength{\tabcolsep}{5pt}
\resizebox{\textwidth}{!}{%
\begin{tabular}{l p{14cm}}
\hline
\rowcolor{gray!15}\textbf{User Engagement - Focused Attention} 
  (\textcolor{ForestGreen}{$F(1,22)=17.888,\ p < 0.005$}) 
 & Q1: I lost myself in this experience 
   (\textcolor{red}{$V=13,\ p = 0.088$}) \\
\rowcolor{gray!15}
 & Q2: The time I spent using the interface just slipped away 
   (\textcolor{ForestGreen}{$V=0,\ p < 0.005$}) \\
\rowcolor{gray!15}
 & Q3: I was absorbed in this experience 
   (\textcolor{ForestGreen}{$V=0,\ p < 0.005$}) \\

\rowcolor{white}\textbf{User Engagement - Perceived Usability} 
  (\textcolor{ForestGreen}{$F(1,22)=9.832,\ p < 0.005$}) 
 & Q4: I felt frustrated while using this interface 
   (\textcolor{ForestGreen}{$V=9,\ p < 0.05$}) \\
\rowcolor{white}
 & Q5: I found this interface confusing to use 
   (\textcolor{red}{$V=13,\ p = 0.297$}) \\
\rowcolor{white}
 & Q6: Using this interface was taxing 
   (\textcolor{ForestGreen}{$F(1,22)=8.609,\ p < 0.05$}) \\

\rowcolor{gray!15}\textbf{User Engagement - Aesthetic Appeal} 
  (\textcolor{ForestGreen}{$F(1,22)=45.195,\ p < 0.005$}) 
 & Q7: This interface was attractive 
   (\textcolor{ForestGreen}{$V=0,\ p < 0.005$}) \\
\rowcolor{gray!15}
 & Q8: This interface was aesthetically appealing 
   (\textcolor{ForestGreen}{$V=0,\ p < 0.005$}) \\
\rowcolor{gray!15}
 & Q9: This interface appealed to my senses 
   (\textcolor{ForestGreen}{$V=0,\ p < 0.005$}) \\

\rowcolor{white}\textbf{User Engagement - Reward Factor} 
  (\textcolor{ForestGreen}{$V=3,\ p < 0.005$}) 
 & Q10: Using the interface was worthwhile 
   (\textcolor{red}{$V=7.5,\ p = 0.074$}) \\
\rowcolor{white}
 & Q11: My experience was rewarding 
   (\textcolor{ForestGreen}{$V=0,\ p < 0.005$}) \\
\rowcolor{white}
 & Q12: I felt interested in this experience 
   (\textcolor{ForestGreen}{$V=0,\ p < 0.005$}) \\

\rowcolor{gray!15}\textbf{Progress} 
  (\textcolor{red}{$V=7.5,\ p = 0.156$}) 
 & Q13: Using this interface, I was able to make good progress toward my goal of gaining interesting insights 
   from the dataset 
   (\textcolor{red}{$V=7.5,\ p = 0.156$}) \\

\rowcolor{white}\textbf{Maintaining Mental Model - Ease of Use} 
  (\textcolor{ForestGreen}{$F(1,22)=18.289,\ p < 0.005$}) 
 & Q14: My interaction with the analysis view is clear and understandable 
   (\textcolor{ForestGreen}{$V=0,\ p < 0.05$}) \\
\rowcolor{white}
 & Q15: Interacting with the analysis view does not require a lot of my mental effort 
   (\textcolor{ForestGreen}{$V=0,\ p < 0.05$}) \\
\rowcolor{white}
 & Q16: I find the analysis view to be easy to use 
   (\textcolor{ForestGreen}{$V=3.5,\ p < 0.005$}) \\
\rowcolor{white}
 & Q17: I find it easy to get the analysis view to do what I want to do 
   (\textcolor{ForestGreen}{$V=0,\ p < 0.005$}) \\

\rowcolor{gray!15}\textbf{Maintaining Mental Model - Usefulness} 
  (\textcolor{ForestGreen}{$V=0,\ p < 0.005$}) 
 & Q18: Using the analysis view in performing the Exploratory Data Analysis task increases my productivity 
   (\textcolor{ForestGreen}{$V=4.5,\ p < 0.05$}) \\
\rowcolor{gray!15}
 & Q19: Using the analysis view enhances my effectiveness in performing the Exploratory Data Analysis task 
   (\textcolor{ForestGreen}{$V=0,\ p < 0.005$}) \\
\rowcolor{gray!15}
 & Q20: Using the analysis view improves my performance in the Exploratory Data Analysis task 
   (\textcolor{ForestGreen}{$V=3,\ p < 0.005$}) \\
\rowcolor{gray!15}
 & Q21: I find the analysis view to be useful in performing the Exploratory Data Analysis task 
   (\textcolor{ForestGreen}{$V=2.5,\ p < 0.05$}) \\
\rowcolor{gray!15}
 & Q22: I find the analysis view to be useful in externalizing the mental model 
   (\textcolor{ForestGreen}{$V=3,\ p < 0.005$}) \\
\rowcolor{gray!15}
 & Q23: I find the analysis view to be useful in understanding the mental model 
   (\textcolor{ForestGreen}{$V=0,\ p < 0.05$}) \\
\rowcolor{gray!15}
 & Q24: I find the analysis view to be useful to perform real-world Exploratory Data Analysis scenarios 
   (\textcolor{ForestGreen}{$V=3,\ p < 0.005$}) \\

\rowcolor{white}\textbf{Maintaining Mental Model - Satisfaction} 
  (\textcolor{ForestGreen}{$V=0,\ p < 0.005$}) 
 & Q25: I am satisfied with the Exploratory Data Analysis process built into this interface 
   (\textcolor{ForestGreen}{$V=0,\ p < 0.005$}) \\
\rowcolor{white}
 & Q26: I am satisfied with the outcome of the analysis using this interface 
   (\textcolor{ForestGreen}{$V=2.5,\ p < 0.05$}) \\

\rowcolor{gray!15}\textbf{Context Selection - Ease of Use} 
  (\textcolor{ForestGreen}{$F(1,22)=36.217,\ p < 0.005$}) 
 & Q27: My interaction with the integrated LLM is clear and understandable 
   (\textcolor{ForestGreen}{$V=0,\ p < 0.005$}) \\
\rowcolor{gray!15}
 & Q28: Interacting with the integrated LLM does not require a lot of my mental effort 
   (\textcolor{ForestGreen}{$V=0,\ p < 0.005$}) \\
\rowcolor{gray!15}
 & Q29: I find the integrated LLM to be easy to use 
   (\textcolor{ForestGreen}{$V=0,\ p < 0.005$}) \\
\rowcolor{gray!15}
 & Q30: I find it easy to get the integrated LLM to do what I want to do 
   (\textcolor{ForestGreen}{$V=0,\ p < 0.005$}) \\
\rowcolor{gray!15}
 & Q31: I find it easy to express my intentions to the integrated LLM 
   (\textcolor{ForestGreen}{$V=0,\ p < 0.005$}) \\

\rowcolor{white}\textbf{Context Selection - Usefulness} 
  (\textcolor{ForestGreen}{$V=4,\ p < 0.05$}) 
 & Q32: Using the integrated LLM improves my performance in the Exploratory Data Analysis task 
   (\textcolor{ForestGreen}{$V=2,\ p < 0.05$}) \\
\rowcolor{white}
 & Q33: Using the integrated LLM in performing the Exploratory Data Analysis task increases my productivity 
   (\textcolor{ForestGreen}{$V=1.5,\ p < 0.05$}) \\
\rowcolor{white}
 & Q34: Using the integrated LLM enhances my effectiveness in performing the Exploratory Data Analysis task 
   (\textcolor{ForestGreen}{$V=2.5,\ p < 0.005$}) \\
\rowcolor{white}
 & Q35: I find the integrated LLM to be useful in performing the Exploratory Data Analysis task 
   (\textcolor{ForestGreen}{$V=1.5,\ p < 0.05$}) \\
\rowcolor{white}
 & Q36: I find the integrated LLM to be useful in performing coding tasks without Python knowledge 
   (\textcolor{red}{$V=8,\ p = 0.109$}) \\

\rowcolor{gray!15}\textbf{Context Selection - Satisfaction} 
  (\textcolor{ForestGreen}{$F(1,22)=27.415,\ p < 0.005$}) 
 & Q37: I am satisfied with the user-LLM interaction mechanism built into this interface 
   (\textcolor{ForestGreen}{$V=0,\ p < 0.005$}) \\
\rowcolor{gray!15}
 & Q38: I am satisfied with the responses provided by the integrated LLM 
   (\textcolor{ForestGreen}{$V=2.5,\ p < 0.005$}) \\
\rowcolor{gray!15}
 & Q39: I am satisfied that the integrated LLM understands my intention 
   (\textcolor{ForestGreen}{$V=1.5,\ p < 0.05$}) \\
\hline
\end{tabular}
}

\end{table*}

\section{User Study}
We conducted a mixed-method user study to explore how \name{}, compared to a baseline system, supports analysts performing real-world EDA tasks, particularly in selecting context for an LLM assistant.
The Baseline consists of a standard JupyterLab environment that uses the traditional linear cell-based layout 
and a built-in LLM chat interface, mimicking Copilot. 
The same LLM model was used for both conditions. 
We chose this baseline because 
it mirrors a typical real-world setup: a standard JupyterLab instance with no special mechanism for organizing cells or content selection assistant for LLM.
Informed by the aforementioned challenges, our study focused on investigating two questions: 
(1) How well does \name{} support analysts to externalize and manage their mental models during EDA with LLM assistants, 
and (2) How well does \name{} help analysts select task-relevant context for LLM assistance in EDA?

\subsection{Participants}
We used the convenience sampling method~\cite{etikan2016comparison} to recruit 12 participants (ten men and two women; aged 20–39) via a local university's mailing lists. 
This group included five PhDs, five Master's, and two undergraduates. 
All participants had experience using both computational notebooks and LLMs for EDA tasks, based on a pre-screening questionnaire. 
Ten participants specialized in Computer Science, one in Mathematics, and one in Accounting.
Their EDA experience ranged from ``a lot'' to ``some,'' with seven reporting either ``some'' or ``quite a bit.'' 
In terms of LLM usage, eight used them ``multiple times a day,'' three ``once a day,'' and one ``several times a week.'' 
Regarding computational notebook experience, eight participants reported using them ``a lot,'' three ``quite a bit,'' three ``a fair amount,'' and one ``some.'' 
Tables~\ref{tab:LLM_for_EDA} and \ref{tab:computational_for_EDA} in \autoref{AppendixA} detail the EDA tasks the participants typically perform with LLMs and computational notebooks. 
The study was approved by the institutional research ethics office.

\subsection{Procedure and Task}
The study was conducted in person over approximately 90 minutes, using a mixed-method design featuring a comparative component, semi-structured interviews, and observational data collection.
Participants were introduced to the study and completed a formal consent process, followed by a demographic questionnaire. 
They then received instructions for two tasks, each completed with a different interface (\name{} or Baseline) within a 15-minute time limit, with some flexibility allowed. Before starting each task, participants underwent a brief training/familiarization session to ensure they were comfortable with the interface.
After each task, participants completed a post-task questionnaire 
capturing their perceptions of subjective factors such as engagement, ease of use, usefulness, and overall satisfaction with the interface.
At the end, participants took part in a semi-structured interview (around 15 minutes) to discuss their experiences and insights. 
The entire session was audio and screen-recorded, and participants received \$30 remuneration.

A Graeco-Latin square design~\cite{martin2005g} was employed to enable a within-subjects~\cite{budiu2018between} comparison of the two interfaces with two distinct tasks. 
Each task was an EDA exercise aimed at uncovering interesting insights from a provided dataset. 
Task \#1, paired with a dataset of billionaires~\cite{kaggleBillionairesDataset}, featured a pre-created notebook with at least 20 cells---representing a long, messy notebook.
Task \#2, using a dataset of hotel bookings~\cite{kaggleHotelBooking}, required participants to begin EDA and externalize their mental model entirely ``from scratch.'' 
This combination of tasks exposed participants to two realistic EDA situations: (1) extending a long, messy, pre-prepared notebook—requiring reconstruction of another analyst’s mental model and more challenging context selection—and (2) starting from scratch, enabling them to build their own mental model and select context grounded in it.
In both interfaces, participants were expected to leverage the integrated LLM to conduct their analyses.

\subsection{Data Collection and Analysis}

To comprehensively compare the systems, we collected four types of data in our study: (1) subjective measures, (2) objective measures, (3) interaction logs, and (4) qualitative feedback. 
These sources enabled us to investigate the effectiveness of the whole interface, the analysis view, and the integrated LLM assistant.

For the overall experience, we assessed user engagement using the User Engagement Scale---Short Form (UES-SF)~\cite{OBrien18}. 
This instrument addresses four dimensions: Focused Attention (FA), Perceived Usability (PU), Aesthetic Appeal (AE), and Reward Factor (RW), each measured with three questions.
For parts related to the mental model support, we measured ease of use, usefulness, and satisfaction. We used a similar approach to measure users' perceptions of context selection support of both systems.
The questions were a combination of selected TAM2~\cite{tam2} and customized questions (see \autoref{tab:subj-questions} for details). 
All subjective measures used 7-point Likert scales, and the responses for each conceptual dimension, including different aspects of user engagement, ease of use, usefulness, and satisfaction)
were averaged into a single score, allowing ordinal data to be analyzed numerically~\cite{lantz2013equidistance, drag_full}. 
We then employed statistical tests to compare the differences between the interfaces, 
using ANOVA~\cite{anova} for normally distributed data and the Wilcoxon Signed-Rank test~\cite{woolson2007wilcoxon} when the data were not normally distributed.
Additionally, we recorded participants' interactions with both interfaces to derive objective measures and visualize user interaction patterns. 
To make sense of the qualitative feedback (see Appendix~\ref{AppendixA} for the questions), we conducted a thematic analysis~\cite{smith_qualitative_2024, thematic}. 
Specifically, we utilized NVIVO~\cite{NVIVO2014} to transcribe participant comments, code them into recurring themes related to each interface (overall interface experience, analysis view, and integrated LLM assistant), and then group these codes to extract high-level insights about participants' experiences.

\begin{figure*}
    \centering
    \includegraphics[width=0.8\linewidth]{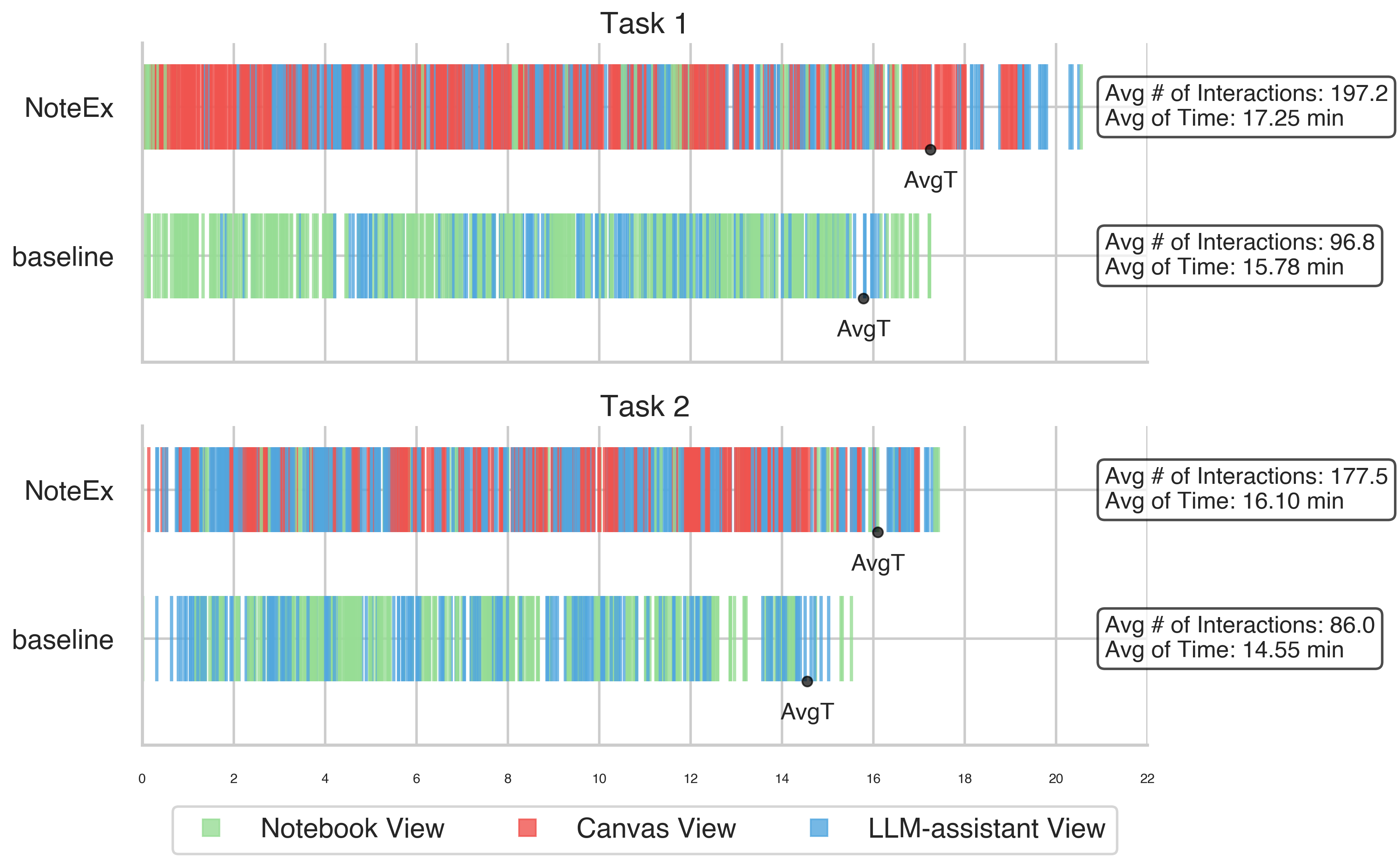}
    \vspace{-4mm}
    \caption{This figure compares interaction logs for Task \#1 (top) and Task \#2 (bottom), and for each interface (Baseline or \name{}) by merging all participants’ interactions into single timelines.  }
    \label{fig:interaction_logs}
\end{figure*}

\begin{table*}[tb]
\centering
\caption{Comparison of key objective metrics, in the format of Mean (Standard Deviation), between Baseline and \name{}. 
}
\vspace{-3mm}
\label{tab:summary-obj-metrics}
\renewcommand{\arraystretch}{1.2}
\setlength{\tabcolsep}{10pt}
\rowcolors{2}{gray!15}{white}
\resizebox{0.7\linewidth}{!}{%
\begin{tabular}{lrr}
\hline
& \textbf{Baseline} & \textbf{\name{}} \\
\hline
Task completion time (min) & 15.17 (1.03) & 16.67 (1.66) \\
\hline
\# of interactions with the Notebook View & 66.25 (42.16) & 16.33 (12.04) \\
\# of interactions with the Canvas View & - & 87.50 (32.30) \\
\# of created cells & 14.67 (8.37) & 15.00 (8.00) \\
\# of total code cells & 10.25 (4.92) & 9.50 (3.99) \\
\# of code cells for plots & 4.25 (2.49) & 5.58 (2.87) \\
\# of code cells for data processing & 6.00 (4.08) & 3.92 (1.66) \\
\# of text/markdown cells & 4.42 (4.54) & 5.50 (4.15) \\
\# of links & - & 14.17 (8.70) \\
\hline
User-LLM interaction time (min) & 1.99 (0.74) & 1.31 (0.67) \\
\# of User-LLM interactions & 25.17 (10.68) & 80.25 (33.19) \\
\# of issued prompts & 6.33 (3.12) & 11.42 (5.07) \\
Prompt length including context (characters) & 9477.82 (3977.66) & 34142.81 (72218.59) \\
Prompt length excluding context (characters) & 1568.75 (1348.86) & 44.26 (19.20) \\
\hline
\end{tabular}
}
\end{table*}

\section{Results}

In this section, we present our findings under three themes regarding participants' overall experience, how they maintain their mental models in EDA, and how they select task-relevant context for LLM assistance. 
Each theme begins by reporting subjective measures and statistical test results, followed by an explanation of why and how,
backed by qualitative feedback, objective metrics (Table~\ref{tab:summary-obj-metrics}), and insights from interaction logs (Figure~\ref{fig:interaction_logs}).

\subsection{General User Experience}

\begin{figure*}[bt]
  \centering
  \includegraphics[width=0.8\linewidth]{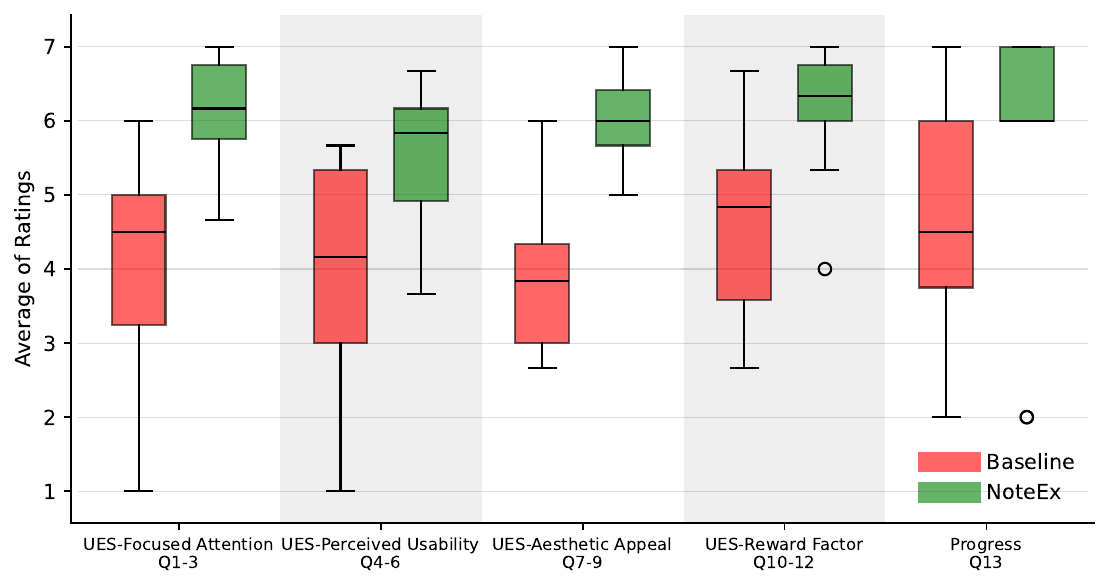}
  \vspace{-4mm}
  \caption{Comparison of overall user engagement metrics (UES-SF) between Baseline and \name{}. 
  }
  \label{fig:quan-subj-general}
\end{figure*}

Overall, participants found \textbf{\name{} superior to Baseline for EDA tasks}. 
As shown in Figures~\ref{fig:quan-subj-general}, \ref{fig:quan-subj-view}, and \ref{fig:quan-subj-llm} and Table~\ref{tab:summary-obj-metrics}, on nearly all grouped metrics (Q1–Q39), participants rated \name{} significantly higher than Baseline, except for Q13 (progress).
Moreover, when asked if they would replace Baseline with \name{}, all participants indicated they would: \pqt{Yes, definitely. Because, as I said, it [\name{}] has reduced my effort and mental load, and it was much easier to use and to visualize my workflow.}{P2}

Participants perceived that \textbf{\name{} was more engaging than Baseline} across four different aspects (Figure~\ref{fig:quan-subj-general}; Table~\ref{tab:subj-questions}:Q1–Q12).
As shown in Table~\ref{tab:summary-obj-metrics}, participants seemed more immersed in the tasks when they used \name{} (an average of 16.67 minutes) compared to using Baseline (an average of 15.17 minutes), while they were instructed to spend 15 minutes for their tasks.
Notably, many participants wished to continue exploring with \name{} even after the allocated time.
In contrast, when using Baseline, several participants indicated they were ready to stop before reaching 15 minutes, and we had to encourage them to continue.
Moreover, participants had almost double of the number of interactions with \name{} compared to that of Baseline in both tasks (Figure~\ref{fig:interaction_logs}), underscoring their greater level of engagement.

By looking into the ratings of Q13 (concerning participants' perception of making good progress toward finding interesting insights; Table~\ref{tab:subj-questions}), Figure~\ref{fig:quan-subj-general} illustrates a few outliers in the \name{} condition, resulting in no significant difference.
However, by removing the two outlier participants, \name{}'s ratings were significantly higher. 
This suggests that \textbf{most participants still perceived \name{} to offer better progress than Baseline.} 
We suspect that the two participants might have been dissatisfied with their findings since the tasks were open-ended; however, further investigation is needed with perhaps more user studies.

Figure~\ref{fig:interaction_logs} visualizes participants' interactions with each interface (\name{} vs. Baseline) across the two tasks.
There emerge four interesting observations.
First, although participants could still use the Notebook View in \name{}, they predominantly relied on the Canvas View (red stripes) instead of the Notebook View (green stripes), suggesting that the Canvas View largely replaced the Notebook View for both interpreting a pre-created mental model (Task \#1) and building a mental model from scratch (Task \#2).
Second, LLM usage (blue stripes) was lower in Task \#1 compared to Task \#2 for both interfaces, likely because, as also observed, participants focused on reviewing existing cells in Task \#1, while in Task \#2 they utilized the LLM more often for code generation.
Third, dense clusters of red stripes in \name{}’s Task \#1 logs reveal that participants spent the initial minutes comprehending the pre-created mental model using the Canvas View, implying that it may have helped establish an understanding of someone else’s mental model.
Finally, \name{} produced nearly twice the total interactions of Baseline, while adding only about two extra minutes on average, indicating higher engagement within a similar overall time frame.

\subsection{Experience with Maintaining Mental Model}

\begin{figure}[tb]
  \centering
  \includegraphics[width=1\linewidth]{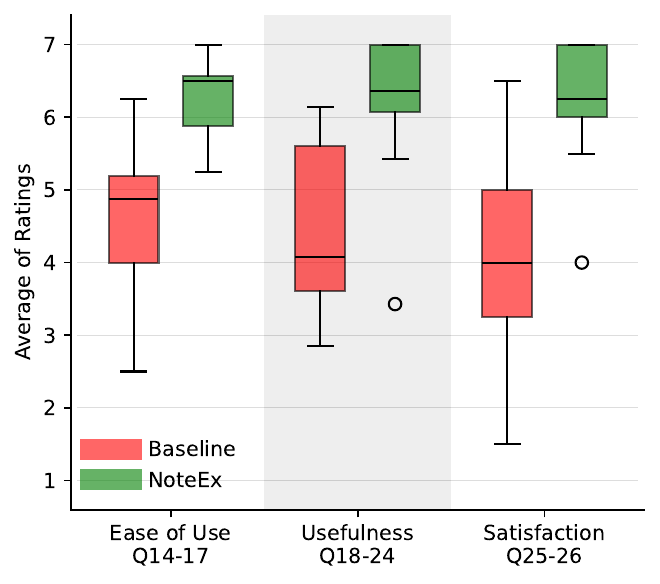}
  \caption{Comparison of user ratings on maintaining mental model between Baseline and \name{}.}
  \label{fig:quan-subj-view}
\end{figure}

As shown in Figure~\ref{fig:quan-subj-view} and Table~\ref{tab:subj-questions}, overall, \textbf{participants considered \name{} to be significantly better than Baseline on maintaining their mental model}, including the ease of use, usefulness, and satisfaction.

\subsubsection{Usability of the Canvas View}
In addition to the Notebook View as in Baseline, \name{} includes the Canvas View that visualizes the mental-model dependencies among cells. 
However, this did not result in a steep learning curve for participants, where they thought \textbf{\name{} was easy and intuitive} (Q14-Q15).
Table~\ref{tab:summary-obj-metrics} indicates that on average 15 cells and 14.17 links were created by participants with \name{}, indicating that they got used to the interface easily.
Also, P8 mentioned that \qt{I really liked the drag-and-drop node system. Adding nodes, deleting nodes, running them, and connecting them.} P3 echoed that \qt{The lines are super helpful because they make it easy to follow my thoughts (mental model).}

Moreover, participants reported that \name{}'s \textbf{Canvas View facilitated navigation through a large number of cells}, which they missed in Baseline. 
\pqt{I liked being able to click on the cells and jump to that location. I thought it was very, very useful.}{P8}. \pqt{When you have many cells, like cell A, then several in between, and cell B, it's useful to quickly navigate back to cell A while adding something to cell B.}{P5}.

\subsubsection{Mental Model Externalization}
Participants found \textbf{\name{} was useful for externalizing their mental model} (Q22).
They pointed out that they could not adequately externalize their mental model using only Baseline's linear Notebook View. 
They needed to specify mental model dependencies to reflect their thinking more clearly: \pqt{The linear structure [in Baseline] made analysis and understanding easier, but it didn't fully align with my mental model. 
Marking relationships across different elements (using the Canvas View) could better reflect how I think about connections in data.}{P8}. 

When asked whether using the Canvas View to externalize their mental model required extra effort, some participants acknowledged that it might be. Nonetheless, they all agreed that \textbf{the manual externalization was worthwhile}, as otherwise they would rely solely on an internal mind map, which increases cognitive load. 
\pqt{Regardless of whether I'm doing the task on the 2D side or 1D side, I still need a mental structure to help me complete the task and understand my current status in the EDA process.}{P1}. 
They viewed the Canvas View as a mind map or an overview of the mental model that complements the Notebook View: \pqt{The right side (Canvas View) gives a big-picture view of the data, showing the overall story, while the left side (Notebook View) focuses on the details of each cell. Using both together makes it easier to understand and work through the analysis.}{P1}.

\subsubsection{Mental Model Understanding} 

Participants emphasized \textbf{the cell-specific information was important in forming an overview of mental model}: \pqt{It's a smoother experience [with \name{}], and it clearly helps me to better see the general picture of how the data interacts with each other.}{P1}. 
P12 noted that color-coded execution statuses helped clarify focus areas: \qt{All the color cues and visuals were a good point. I also expected to click that myself.}
P3 appreciated the ability to view the output of all visualizations at once: \qt{With the 1D view, you can see only one type of visualization at a time. 
But with this (Canvas View), you can see all the visualizations simultaneously on a mind map.} 
They also valued the Data Information View, particularly for seeing where data is defined and used as well as understanding the data structure: \pqt{Instead of searching through the code [in Baseline], you can quickly see where a variable is defined and used, with previews available.}{P3}.
However, one limitation they noted was that the view only shows the schema, not the actual data. To address this, they often had to open the raw CSV file. Some participants suggested that the Data Information View could be enhanced with a data preview or simple visualization: \pqt{Maybe users could click on it to see the actual data for that parameter. Or perhaps there could be a very basic, more general visualization of that parameter instead.}{P12}

In Task \#1, participants had to understand a pre-created notebook; they reported that \name{} enabled them to more easily comprehend not only their own mental model but also others' (Q23). 
As shown in Figure~\ref{fig:interaction_logs}, during both tasks, participants who had access to the Canvas View relied significantly less on the Notebook View, despite its availability. 
Especially at the beginning of Task \#1, participants depended heavily on the Canvas View to grasp another analyst's mental model. 
They noted that the \textbf{Canvas View helped them understand others' mental model more efficiently}: \pqt{If I had just come in here and had no context of what the dataset was, it would give a good overview of what the previous user was looking at.}{P3}. 
They also thought it saved considerable time: \pqt{The mind map (Canvas View) made it much easier to understand your workflow and intent quickly. Without it [in Baseline], I would've spent 10-15 minutes just figuring out your approach, but with the cell structure, I understood it instantly.}{P4}.

\subsubsection{EDA Performance}
Overall, \textbf{\name{} was considered more useful in conducting EDA than Baseline} because participants thought their performance was enhanced (Q18-21), especially for real-world (Q24) and complex tasks. 
\pqt{If we had a more demanding task that required greater mental attention, knowing the relationships between cells in a graph would be more helpful than for a simple task.}{P5}.
As shown in Table~\ref{tab:summary-obj-metrics}, comparing Baseline vs. \name{}, although the total number of created cells remained similar (14.67 vs. 15.00), participants made greater use of text/markdown cells (4.42 vs. 5.50) and dedicated more cells to plotting (4.25 vs. 5.58). 
This potentially indicates that the Canvas View facilitated more thorough documentation and visualization, further contributing to enhanced performance in complex scenarios.

\textbf{Participants appreciated the selective or path-based execution in \name{} for supporting EDA}. 
First, many stressed saving time by avoiding unnecessary re-runs (P1, P3, P6, P8). 
This approach allowed them to skip already-executed cells and focus only on the tasks at hand: \pqt{You're only rerunning the cells when you need to. And I think that (path-based execution) could save a lot of resources and time.}{P3}. 
Second, participants highlighted the convenience of consolidating multiple analyses into a single notebook, rather than spreading them across multiple files, which streamlined their workflows and mental models: \pqt{I used to just break down my analysis into different files. So now, with that feature (Canvas View) I can use, I can do all of my analysis on a certain data set on one file.}{P2}. 
Finally, participants noted potential memory and performance improvements through path-based execution: \pqt{It always seems like the linear notebook [in Baseline] is very memory inefficient compared to what this (\name{}) could be, because here you have a dependency graph.}{P12}.
While path-based execution is not directly related to context selection, it implicitly supported participants in maintaining a coherent mental model by allowing them to focus on relevant information.

\subsection{Experience with Selecting LLM Context}

\begin{figure}[tb]
  \centering
  \includegraphics[width=1\linewidth]{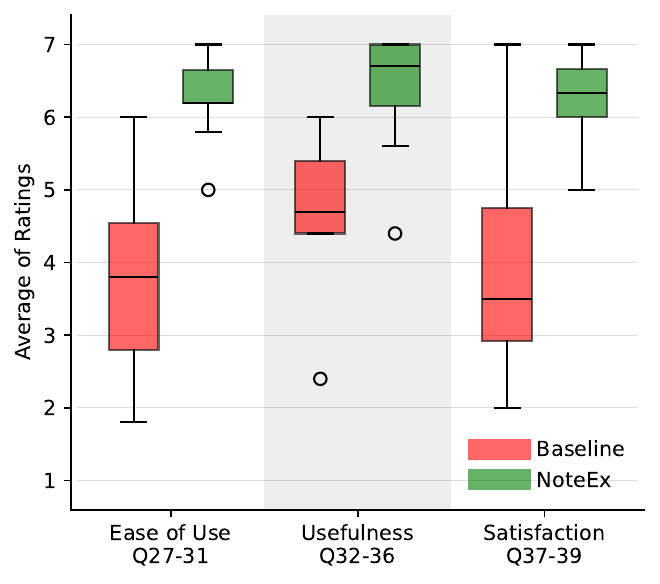}
  \caption{Comparison of user ratings on context selection between Baseline and \name{}.}
  \label{fig:quan-subj-llm}

\end{figure}

Overall, as shown in Figure~\ref{fig:quan-subj-llm} and Table~\ref{tab:subj-questions}, participants found their \textbf{experience with the LLM assistant in \name{} was significantly better than Baseline} on all three aspects, including ease of use, usefulness, and satisfaction.

\subsubsection{Interactive Context Selection and Prompting Efficiency}
Participants found their tasks easier because \textbf{the LLM assistant of \name{} offered more flexible selection of context by interactivity}, eliminating the need for repetitive copying and pasting (Q27-Q28): \pqt{I think the first one (\name{}) is better in a way because I can replace it directly. I didn't have to copy and paste; I could just select the context and replace it.}{P11}. 
Also, they noticed less manual typing was required, reducing the burden of extensive text-based instructions: \pqt{Because it (\name{}) knew my context, it had an understanding of my data, and I didn't have to type much into the system. I just had to press a few buttons.}{P2}. 
As shown in Table~\ref{tab:summary-obj-metrics}, for Baseline vs. \name{}, the pronounced difference in Prompt Length/Characters (1568.75 vs. 44.26 characters) highlights that participants with Baseline often had to type lengthy prompts to convey their intentions. 
Additionally, fewer clarifications needed may indicate that \textbf{the LLM assistant quickly grasped users' intention} (Q31): \pqt{It (\name{}) really understood my intentions really well without me having to type everything.}{P2}. 
\textbf{Participants also highlighted reduced mental effort with \name{}} throughout the process, making exploration and debugging simpler (Q28).

\subsubsection{EDA and Programming Support}
The LLM-assistant View in \name{} was considered useful in terms of enhancing their performance (Q32-Q35). 
By providing immediate, contextually appropriate code suggestions, \textbf{participants thought \name{} reduced coding friction}---the extra time and effort spent debugging, rewriting, or clarifying code before achieving desired outcomes: \pqt{The second one (\name{}) surprisingly got a lot of the code right the 1st time around. It made sense of the context pretty well as well.}{P12}. 
P12 further commented that it allowed them to concentrate on higher-level insights by minimizing cognitive load. 
Table~\ref{tab:summary-obj-metrics} confirms these benefits: from Baseline to \name{}, on average, the time of user-LLM interaction dropped from 1.99 to 1.31 minutes, while the total number of user-LLM interactions rose from 25.17 to 80.25 and the number of issued prompts increased from 6.33 to 11.42.

Participants also reported different activities with which the integrated LLM in \name{} helped them. 
For coding, participants mentioned \textbf{they could quickly generate and refine snippets in \name{}} without writing everything from scratch, as P9 noted: \qt{I found it (\name{}) more useful in coding tasks, because I don't have to write the code out myself. So it's very easy to go and replace and stuff.} 
For debugging, participants highlighted that \textbf{troubleshooting in \name{} was simplified because they did not always need to supply extensive context}: \pqt{It's also easier to debug, because sometimes I don't need to tell the LLM that much information.}{P1}. 
In contrast, participants noted that Baseline required repeatedly copy-pasting error messages and code, which needs extra effort:
\pqt{Every single time I need to include extra explanations or copy-paste the errors, and that takes effort.}{P1}.
Finally, for explaining outputs, \textbf{participants appreciated receiving immediate interpretations of plots or tables}: \pqt{I found that (LLM-assistant of \name{}) pretty convenient because I'm generating a plot that isn't from me, so I can quickly see what it's about.}{P1}.
In comparison, Baseline offered no direct way to attach visual outputs to the LLM: \pqt{In the Denmark example, I could quickly ask, `Hey, why is there no Denmark information in this graph?' You can’t really do that in the regular Jupyter Notebook linear thing.}{P8}.

\textbf{Several participants saw the integrated LLM as beneficial for novice programmers}, as it reduced or even removed the need to write code manually (Table~\ref{tab:subj-questions}-Q36), while there was no significant difference between the two conditions. 
\pqt{The second one (\name{}), I would say, didn't feel like programming at all, because it was basically just thinking about what I saw in the data.}{P12}. 
In contrast, the Baseline interface was said to pose a higher barrier for novices, since it demanded more direct coding or domain knowledge: \pqt{With the first one (Baseline), you do need some kind of domain knowledge. I’m not too familiar with pandas and numpy, and all of those libraries.}{P3}.
Also, a few participants voiced the concern that without any coding background, one might not catch LLM hallucinations, leading to misleading outputs: \pqt{Those who do not have any Python knowledge will think that whatever AI is generating is correct.}{P11}. 

\subsubsection{LLM Task Performance}
One main reason that participants felt their performance increased was that \textbf{they perceived the integrated LLM of \name{} well understood their intention} (Q39). 
First, participants found that they did not need to repeatedly explain what they wanted: \pqt{It understood me on what I was meaning, because I didn't have to explain myself too much.}{P2}. 
Second, they saw the interaction as more conversational and context-aware, which streamlined their workflow; for example, P3 stated, \qt{It's nicer in a way, because it already knows the dataset and it feels like you're having a conversation where both people understand what you're talking about.}{P3}.

\textbf{Participants repeatedly highlighted that manually providing proper context caused LLMs to understand them better}, whereas, with Baseline, the LLMs could be easily overwhelmed without focusing on the most relevant information. 
P10 explained, \pqt{Giving too much context [as in Baseline] confused the LLM and could make it hallucinate more.}{P10}, emphasizing the risk of simply dumping everything into the prompt. 
P3 also felt that \pqt{Instead of like just giving it this whole CSV every time, it's really nice to just pick this and that,}{P3} noting how interactively selecting the data or code reduced both confusion and effort. 
Although manually providing context in \name{} may introduce some extra effort, \textbf{participants appreciated that the system leveraged links between nodes to make mental-model-based default context selections} whenever they opened the integrated LLM. 
They also valued the flexibility to revise these suggestions when necessary: \pqt{Because it's connected it automatically selected, then you can add it.}{P1}

It should be noted that the \textbf{participants' ability to verify and/or select the proper context was closely tied to keeping track of and understanding of the mental model} developed using the Canvas View. 
\pqt{I knew it was ChatGPT, but its (the Canvas View's) ability to get context and relate to cells and graphs made me more confident in its answers, so I knew it had the right information.}{P8}
\pqt{keeping track of what context might be required can sometimes be overwhelming if you have, like a spread out mental model of it.}{P12}
Moreover, the integrated environment allowed them to interact with the LLM without repeatedly providing low-level details: \pqt{I could easily give instructions without specifying the variable or cell.}{P5} 
In contrast, participants pointed out that Baseline required explicitly providing all these details every time, making it a more cumbersome experience:
\pqt{But when I was working with the second interface (Baseline), I had to specify all this, and that’s time-consuming.}{P5}

\section{Discussion}

As indicated in our design and evaluation, \name{} supports analysts maintaining their mental model through its Canvas View, a flowchart-style visualization that enables efficient navigation across many cells, externalizes evolving EDA progresses (addressing CH1), surfaces cell-level details and data information level for an at-a-glance overview (addressing CH2 and CH3), and accelerates understanding of mental model.  
Its context selection mechanism flexibly aligns responses with user intent, lowering cognitive and coding effort, and provides mental model-informed default contexts to mitigate manual-selection overhead, while letting context verification and refinement (CH4).
In the following subsections, we first discuss the implications for future work, then the limitations and potential improvements.

\subsection{Design Implications}
\label{implications}

Our findings have several implications for LLM-assisted programming beyond using notebooks. 
In many IDE workflows (e.g., React, Python), long and fragmented codebases make automatic context selection error-prone and resource-intensive: agents may select a the wrong file or function simply because it is open, or expend costly tokens and latency to discover the right context. 
Learning from our study, a human-in-the-loop context selection mechanism on top of automatic methods could reduce both error and cost.
To support such a mechanism, future systems could surface interactive codebase overviews (e.g., call/data-flow graphs, feature- or task-scoped subgraphs, and recent-edit heatmaps), enabling users to curate precise, task-relevant slices of code that are passed to the model. 
Such designs raise concrete research questions about trade-offs between fully automatic retrieval and guided manual selection, including impacts on accuracy, turnaround time, token usage, and user trust.
Evaluations should compare hybrid strategies that blend lightweight automatic proposals with user-approved, visual selection to balance efficiency with control.

A second implication concerns the representational form of notebooks. 
Consistent with prior work~\cite{Harden2023, Harden2022Exploring_Organization, christman20232d}, our results suggest that while traditional notebooks are presented linearly (1D), analysts’ mental models of analysis workflows are inherently spatial and relational (2D). 
Echoing a participant’s remark---\pqt{It didn’t complement it. It just totally replaced it.}{P2}---usage patterns showed that the Canvas View often supplanted the Notebook View (Figure~\ref{fig:interaction_logs}). 
This invites the provocative question of whether a 2D representation could wholly replace the 1D notebook. 
As future directions, tools can be designed only with a 2D workspace, but thoughtfully integrates classic 1D affordances, such as fine-grained code editing and detailed outputs, directly into the 2D canvas.
Such a design would require careful interaction techniques for editing, selection, debugging, and versioning within a spatial layout. 
A controlled study should test whether a 2D-only approach maintains or exceeds performance, learnability, and satisfaction relative to mixed or 1D-only baselines.

Additionally, flowcharts produced by participants after completing their analysis tasks (\eg P1’s task 2; see Appendix~\ref{AppendixA}) verified that analysts relied on multiple types of mental model dependencies: sequential exploration (next step), parallel exploration (alternate paths), hierarchical stages (parent to child), fork-and-join (diverging from a cell or converging to a cell), and documentation links. These patterns, observed in \name{}, underscore the importance of mental model dependencies in EDA and their difference with code dependencies.
Existing EDA tools typically encode and leverage code dependencies (e.g., dataflow, execution order)~\cite{Harden2023, Rule2018, Head2019, MOON2023}, but overlook the analyst’s higher-level, task-driven relations. 
Future notebook systems can incorporate both layers: use code dependencies for correctness and reproducibility, and mental-model dependencies for organization, navigation, explanation, and collaboration. 
Concretely, tools might support authoring and maintaining mental-model links; infer candidate links from interaction traces; and use them to drive layout, summarization, and retrieval for LLM assistance. 
A key direction is to evaluate whether making mental-model dependencies first-class improves comprehension, refactoring, onboarding, collaboration, and data storytelling compared to code-only dependency views.

\subsection{Limitations and Future Work}

While \name{} offers substantial improvements over traditional computational notebooks, several limitations should be acknowledged. 
First, visual scalability may become an issue for very large notebooks. As cells grow, the flowchart can become crowded, making it difficult to navigate or spot relevant connections. Although \name{} includes a minimap, users might struggle with very large analyses.
To address this, future versions of \name{} could introduce collapsible sub-graphs or clustering for related cells, as suggested by P2: \qt{Clustering those visualizations or code blocks can be a helpful thing for the mental model.} 
Combined with the minimap, this could further streamline navigation in sprawling workflows, keeping \name{} manageable even when the number of cells grows large.

Second, \name{} relies on users to manually define dependencies between cells in the Canvas View. 
While effective and flexible, it can be time-consuming and laborious. 
We partially mitigated this with drag-and-drop actions that create and link a cell in one single interaction. 
However, when users first create cells in the linear (1D) notebook view and then switch to the canvas to add links, they incur additional steps. 
Possible solutions include integrating mechanisms to motivate the analyst utilize the canvas or retire the traditional 1D view and rely exclusively on the Canvas View. 
As noted in Section~\ref{implications}, this would require careful redesign of onboarding and discoverability, a clear migration path for existing workflows, and rigorous user evaluation.

Third, manual dependency specification might be error-prone. 
Inaccurate links (i.e., failures to externalize mental-model dependencies correctly) can yield incorrect context selection and lower-quality responses. 
Automating aspects of dependency management might be essential~\cite{Milo2018}. 
Future tools could leverage LLMs to learn from users’ prior mental models---as captured in Canvas Views from past files---by analyzing patterns across notebooks to identify common workflows and dependency conventions. 
The tool could automatically suggest, or generate, an initial Canvas View for a new notebook that aligns with the user’s mental model. 
This personalized automation might reduce user burden; however, its effectiveness should be systematically studied. 
Beyond dependency management, more advanced LLM capabilities could further enhance such tools~\cite{ma2023demonstrationinsightpilotllmempoweredautomated}, for example by automatically generating concise summaries of the entire analysis, including progress and findings, and by offering proactive suggestions that surface potential issues or highlight unexplored analysis paths. 
Such suggestions may bias or overconstrain the analyst’s mental model and risk premature convergence; thus, any automation should be optional, transparent, and evaluated for unintended guidance effects.

Lastly, our user study was conducted with a relatively small sample size of 12 participants, all of whom had prior experience with computational notebooks and LLMs. 
The findings are promising, however, they may not fully generalize to a broader population of data analysts with varying levels of expertise or to different types of data analysis tasks. 
Conducting larger-scale studies with diverse user groups and datasets would help validate and generalize the findings from our user study. 
Such studies could also explore the long-term usability and adoption of \name{} in real-world data analysis scenarios, providing insights into how the system performs over extended periods and in different contexts.
Furthermore, the study used a specific LLM (GPT-4o-mini), and the performance of \name{} might vary with different models or as LLMs continue to evolve.

\section{Conclusion}

In this work, we conducted a formative study to characterize the challenges analysts face when selecting task-relevant context for LLM-assisted EDA in computational notebooks. 
Guided by the identified four main challenges, we introduced \name{}, a JupyterLab extension that externalizes analysts’ mental models and leverages them for context selection. 
\name{} combines a Canvas View for visualizing mental-model dependencies and cell metadata, a Data Information View that surfaces variables and their provenance, and an LLM-Assistant View that proposes editable, task-relevant context based on mental model dependencies.
A user study with 12 participants demonstrated that \name{} improved engagement, ease of use, usefulness, and satisfaction over a baseline, while better supporting mental model maintenance and precise context selection. 
Participants crafted shorter prompts, needed fewer clarifications for LLM, and received higher-quality LLM responses.
Our findings point to critical implications for context selection: pair automatic retrieval with human-in-the-loop, visually guided selection; elevate mental-model dependencies to first-class alongside code dependencies; and adopt 2D, task-centric overviews to curate minimal, relevant context. 
Such hybrids can improve accuracy, reduce hallucinations and token costs, and strengthen user trust.

\bibliographystyle{ACM-Reference-Format}
\bibliography{references.bib}

\clearpage
\appendix

\section{Prompt engineering}
\label{AppendixI}

Below, we illustrate how prompts, including both the system instruction and user message, are dynamically crafted based on different contextual parameters.

\subsection*{Parameters}
\begin{itemize}
    \item \textbf{promptType}: one of \verb|fix|, \verb|modify_code|, \verb|gen_code|, or \verb|explain|
    \item \textbf{fullText}: the entire code in the cell (could be empty)
    \item \textbf{selectedText}: any highlighted fragment (could be empty)
    \item \textbf{prompt}: the user's request or instructions
    \item \textbf{contextCell}: contents of related context cell(s) (if any)
    \item \textbf{contextData}: data variables/arguments and sample structures (if any)
    \item \textbf{checkBox\_error}: (If checked and debugging was selected as promptType) an error string 
    \item \textbf{checkBox\_output}: (If checked) output(s) of the cell that can be image, table, or text.
    \item \textbf{checkBox\_condense}: (If checked) instructs a concise response
\end{itemize}

\subsection{System Instructions}

Each box shows how the system instruction is formed based on \texttt{promptType}. Text in brackets \verb|[...]| appears only if that parameter is non-empty. Bullets indicate what to do with \textit{selectedText} and \textit{fullText}.

\vspace{1em}
\noindent
\begin{tcolorbox}[
    colback=gray!10,
    colframe=gray!50!black,
    fontupper=\ttfamily\footnotesize,
    breakable,
    title=System instruction for debugging \\ (promptType: fix)
]
You are an AI assistant that modifies Python code to fix runtime errors in a Jupyter Notebook cell.\\
You receive:\\
-- [the full code if \textit{fullText} is non-empty]\\
-- [the selected part of the code if \textit{selectedText} is non-empty]\\
-- [the context cell(s) if \textit{contextCell} is non-empty]\\
-- [the context data(s) if \textit{contextData} is non-empty]\\
-- [the outputs if \textit{checkBox\_output} is non-empty]\\
-- the error message (runtime error)\\
-- a prompt (the user’s instructions)

$\bullet$ If \textit{selectedText} is non-empty, only fix that portion; do not alter other parts. Output the \emph{full corrected code} that replaces the selected portion.\\
$\bullet$ If \textit{selectedText} is empty but \textit{fullText} is non-empty, output the \emph{entire corrected code}.

At the beginning of the code, include:
\texttt{"\#\#\# Generated by AI at [timestamp]; Prompt: [prompt]"}
as a comment. Do not remove existing lines starting with \texttt{"\#\#\# Generated by AI"} from the input.
\end{tcolorbox}

\noindent
\begin{tcolorbox}[
    colback=gray!10,
    colframe=gray!50!black,
    fontupper=\ttfamily\footnotesize,
    breakable,
    title=System instruction for modifying existing code \\ (promptType: modify\_code)
]
You are an AI assistant that modifies Python code in a Jupyter Notebook cell.\\
You receive:\\
-- [the full code if \textit{fullText} is non-empty]\\
-- [the selected part of the code if \textit{selectedText} is non-empty]\\
-- [the context cell(s) if \textit{contextCell} is non-empty]\\
-- [the context data(s) if \textit{contextData} is non-empty]\\
-- [the outputs if \textit{checkBox\_output} is non-empty]\\
-- a prompt (the user’s instructions)

$\bullet$ If \textit{selectedText} is non-empty, modify only that portion; do not alter unselected parts. Output \emph{only the new content} that replaces the selection.\\
$\bullet$ If \textit{selectedText} is empty but \textit{fullText} is non-empty, output the \emph{entire modified code}.

At the beginning of the code, include:
\texttt{"\#\#\# Generated by AI at [timestamp]; Prompt: [prompt]"}
as a comment. Do not remove existing lines starting with \texttt{"\#\#\# Generated by AI"} from the input.
\end{tcolorbox}

\noindent
\begin{tcolorbox}[
    colback=gray!10,
    colframe=gray!50!black,
    fontupper=\ttfamily\footnotesize,
    breakable,
    title=System instruction for generating new code \\ (promptType: gen\_code)
]
You are an AI assistant that generates Python code for a Jupyter Notebook cell.\\
You receive:\\
-- [the full code if \textit{fullText} is non-empty]\\
-- [the context cell(s) if \textit{contextCell} is non-empty]\\
-- [the context data(s) if \textit{contextData} is non-empty]\\
-- [the outputs if \textit{checkBox\_output} is non-empty]\\
-- a prompt (the user’s instructions)

$\bullet$ If \textit{fullText} is non-empty, your output should be \emph{the new code} appended to the existing code.

At the beginning of the code, include:
\texttt{"\#\#\# Generated by AI at [timestamp]; Prompt: [prompt]"}
as a comment. Do not remove existing lines starting with \texttt{"\#\#\# Generated by AI"} from the input.
\end{tcolorbox}

\noindent
\begin{tcolorbox}[
    colback=gray!10,
    colframe=gray!50!black,
    fontupper=\ttfamily\footnotesize,
    breakable,
    title=System instruction for explaining code/output or answering questions \\ (promptType: explain)
]
You are an AI assistant that explains code/output or answers questions about them.\\
You receive:\\
-- [the full code if \textit{fullText} is non-empty]\\
-- [the selected part if \textit{selectedText} is non-empty]\\
-- [the context cell(s) if \textit{contextCell} is non-empty]\\
-- [the context data(s) if \textit{contextData} is non-empty]\\
-- [the outputs if \textit{checkBox\_output} is non-empty]\\
-- a prompt (the user’s instructions)

$\bullet$ If \textit{selectedText} is non-empty, focus on that portion while considering the full code/context.\\
$\bullet$ If \textit{checkBox\_condense} is true, produce a concise response (max three sentences).

At the beginning of your response, include:
\texttt{"\#\#\# Generated by AI at [timestamp]; Prompt: [prompt]"}
Do not remove existing lines starting with \texttt{"\#\#\# Generated by AI"} from the input.
\end{tcolorbox}

\subsection{User Message Templates}

After the system instruction, we assemble a single \texttt{user} message. Any lines in square brackets appear only if that parameter is non-empty or relevant:

\begin{tcolorbox}[
    colback=gray!10,
    colframe=gray!50!black,
    fontupper=\ttfamily\footnotesize,
    breakable,
    title=User message template
]
[If \textit{fullText} is non-empty:\\
\quad ``The full code: \textit{fullText}'']\\

[If \textit{selectedText} is non-empty:\\
\quad ``\textbf{***} The selected part: \textit{selectedText}'']\\

[If \textit{promptType} starts with ``fix'':\\
\quad ``\textbf{***} the error I got after executing the code: \textit{checkBox\_error}'']\\

[If \textit{contextCell} is non-empty:\\
\quad ``\textbf{***} the context cell(s): \textit{contextCell}'']\\

[If \textit{contextData} is non-empty:\\
\quad ``\textbf{***} the context data(s): \textit{contextData}'']\\

[If \textit{checkBox\_output} is non-empty:\\
\quad ``\textbf{***} the output(s): (parsed text or 'attached as image(s)')'']\\

``\textbf{***} the prompt: \textit{prompt}''
\end{tcolorbox}

\section{Comparative study details}
\label{AppendixA}

\subsection{Participant Characteristics}

\begin{table}[h!]
\centering
\small
\caption{Typical LLM usage for EDA tasks by participants.}
\vspace{-4mm}
\label{tab:LLM_for_EDA}
    \begin{tabular}{lll}\toprule
    \textbf{{Task Category}} & \textbf{{Sub-category}} & \textbf{{\# of Participants}}\\\midrule
    Coding & Generally & 4 \\
     & Debugging & 3 \\
     & Generation & 2 \\
     & Refactoring & 1 \\\midrule
    Data preprocessing &  & 2 \\\midrule
    Data visualization &  & 2 \\\midrule
    Machine Learning &  & 1 \\\bottomrule
    \end{tabular}
\vspace{1mm}
\end{table}

\begin{table}[h!]
\centering
\small
\caption{Typical computational notebook usage for EDA tasks by participants.}
\vspace{-4mm}
\label{tab:computational_for_EDA}
    \begin{tabular}{lll}\toprule
    \textbf{{Task Category}} & \textbf{{\# of Participants}}\\\midrule
    Data visualization & 5 \\\midrule
    Machine Learning & 5 \\\midrule
    Identifying pattern in data & 4 \\\midrule
    Data preprocessing & 3 \\\midrule
    Qualitative analysis & 2 \\\midrule
    Quantitative analysis & 1 \\\midrule
    Medical imaging & 1 \\\midrule
    Quantum computing & 1 \\\bottomrule
    \end{tabular}
\vspace{1mm}
\end{table}

\subsection{Interview Questions}
Not all of the questions were asked. The questions were picked based on observations of users' interactions with the systems.

\begin{enumerate}
    \item \textbf{General Experience}
    \begin{enumerate}[label*=\arabic*.]
        \item How would you describe your overall experience using [\name{}] compared to traditional Jupyter notebooks?
        \item How intuitive was the interface of \name{}? Were there any features that were particularly easy or difficult to use?
        \item Which features of \name{} did you find most and least beneficial, and why?
    \end{enumerate}
    
    \item \textbf{RQ1: 2D Canvas and Externalization}
    \begin{enumerate}[label*=\arabic*.]
        \item How did the 2D canvas in \name{} help or didn’t help you externalize your mental model of the EDA task? Please explain why it did or didn’t.
        \item How did the 2D canvas in \name{} impact your understanding of the EDA task? Did it affect your performance? If so, how?
        \item How helpful or unhelpful was the contextual information shown in the canvas, such as data variables, cell content snippets, execution status, or output previews, for performing the task? Please explain.
        \item How helpful or unhelpful was the data variable view when performing the task? Please explain.
        \item How did you find the ability to execute cells using the path? Please elaborate.
        \item Can you share a specific instance where the 2D view helped you accomplish something more easily or effectively than traditional notebooks?
    \end{enumerate}
    
    \item \textbf{RQ2: LLM Integration}
    \begin{enumerate}[label*=\arabic*.]
        \item How effective was the integrated LLM in assisting with tasks such as code explanation, debugging, or generation? Please explain why it was or wasn’t helpful.
        \item For which specific tasks (e.g., output analysis, code explanation, debugging, or generation) did the LLM feel most useful, and why?
        \item Can you describe a specific instance where you utilized the LLM?
        \item How did selecting contextual information (e.g., variables, dependencies) for the LLM affect the relevance of its responses? Please explain.
        \item Did the LLM’s awareness of cell dependencies or variable usage increase or decrease your trust in its responses? Why or why not?
        \item How well did the integrated LLM in [\name{}] understand your intent? What aspects contributed to this perception?
        \item Can you rely on the LLM if you have no Python experience? Compare baseline vs. \name{}.
    \end{enumerate}

    \item \textbf{Final Questions}
    \begin{enumerate}[label*=\arabic*.]
        \item What improvements would you suggest for the system?
        \item Do you see yourself replacing traditional notebooks with \name{}? Why or why not?
        \item Any final thoughts?
    \end{enumerate}
\end{enumerate}

\subsection{An example of resulting flowchart}

\begin{figure*}
    \centering
    \includegraphics[width=\linewidth]{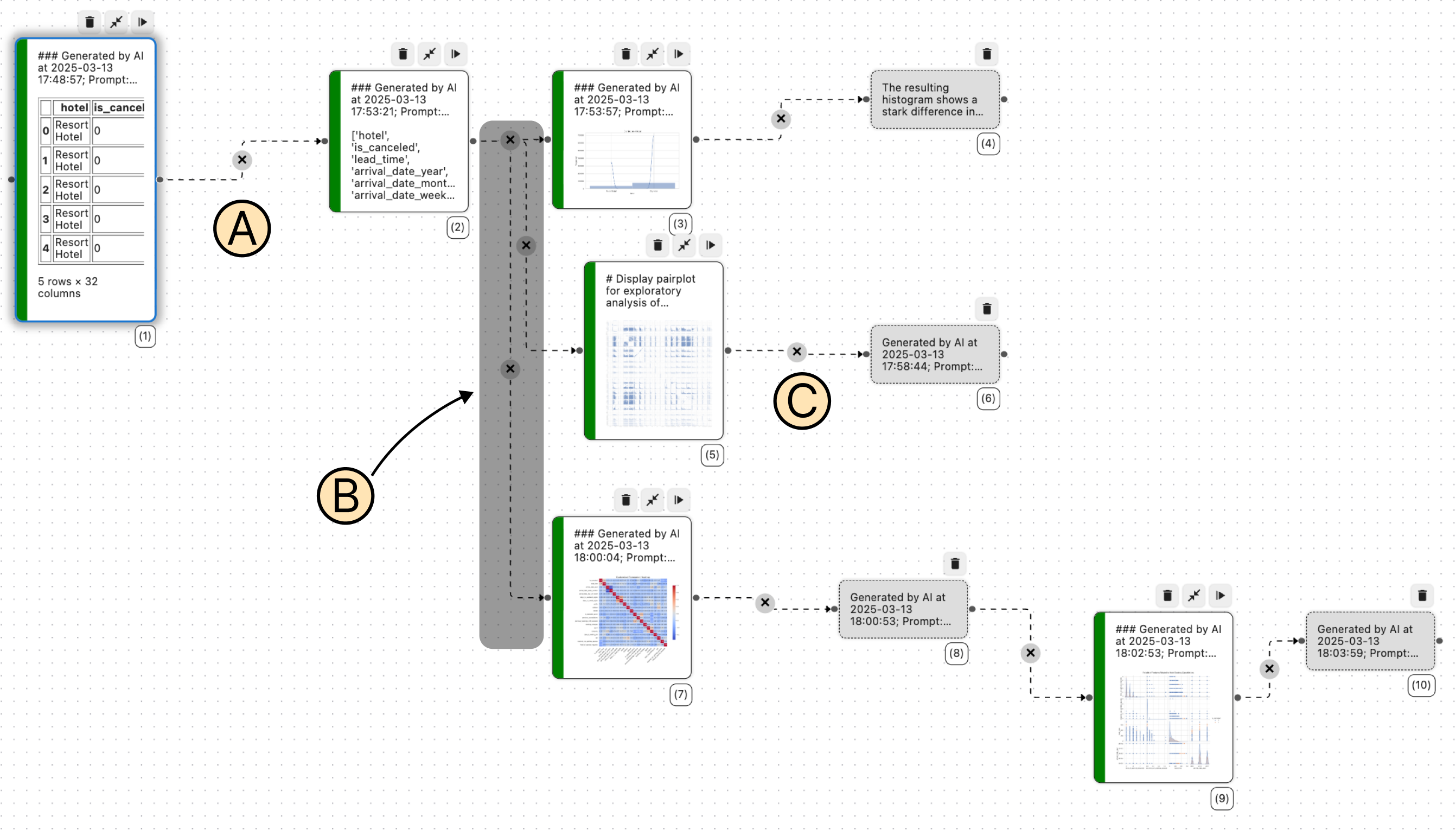}
    \vspace{-4mm}
    \caption{An example flowchart created by P1 after completing Task 2, which involved performing the analysis from scratch. (A) Sequential Exploration (B) fork (C) Documentation}
    \label{fig:flowchart_example_task2_p1}
\end{figure*}

\end{document}